\documentclass[12pt]{article}
\usepackage[utf8]{inputenc}
\usepackage{xcolor}
\definecolor{darkblue}{rgb}{0, 0, 0.699219}
\definecolor{darkgreen}{rgb}{0, 0.699219, 0}
\definecolor{darkred}{rgb}{0.699219, 0, 0}
\definecolor{purple}{rgb}{0.5, 0, 0.25}
\definecolor{webbrown}{rgb}{0.601562, 0, 0}
\definecolor{webgreen}{rgb}{0, 0.398438, 0}
\usepackage{amsmath}
\usepackage{amsthm}
\usepackage{amssymb}
\usepackage{graphicx}
\usepackage{setspace}
\usepackage[authoryear]{natbib}
\usepackage[bookmarks=false,breaklinks=false,pdfborder={0 0 1},colorlinks=false]{hyperref}
\hypersetup{colorlinks,citecolor=darkred,filecolor=black,linkcolor=darkblue,urlcolor=webgreen,pdfstartview=FitH}
\hypersetup{pagebackref=false}

\makeatletter
\usepackage{amsfonts}
\usepackage{caption}
\usepackage{soul}% used for striking out sentences.

\usepackage{comment}
\usepackage{color}
\usepackage{fullpage}

\newtheorem{lemma}{{\bf Lemma}}\newtheorem{proposition}{{\bf Proposition}}\newtheorem{corollary}{{\bf Corollary}}\newtheorem{theorem}{{\bf Theorem}}\newtheorem{defn}{{\bf Definition}}\newtheorem{obs}{{\bf Observation}}\newtheorem{claim}{{\bf Claim}}\newtheorem{example}{{\bf Example}}\newtheorem{remark}{{\bf Remark}}

\allowdisplaybreaks

\title{{\bf von Neumann-Morgenstern stable sets \\ in job matching problems}\thanks{We are grateful to Jean-Jacques Herings and participants at a workshop on matching at Indian Statistical Institute, Delhi for comments and feedback. \texttt{refine.ink} was used to check consistency of proofs and exposition of this article.  Debasis Mishra acknowledges financial support from ANRF through Project ANRF/ARGM/2025/000572/QSS.}}
\author{Madhuparna Karmokar and Debasis Mishra~\thanks{Karmokar: Indian Institute of Management Calcutta, \texttt{madhuparnak@iimcal.ac.in}. Mishra: Indian Statistical Institute, Delhi, \texttt{dmishra@isid.ac.in}.}}

\makeatother

\begin{document}

\allowdisplaybreaks 

\maketitle 

\begin{abstract}
We show that a von Neumann-Morgenstern (vNM) stable set generically exists in the many-to-one job-matching model. Unlike the core, which can be empty in this model, a vNM stable set exists whenever there is a unique efficient matching: the set of all individually rational (IR) and efficient outcomes form a vNM stable set. If an economy has multiple efficient matchings, a vNM stable set may not exist. If the vNM stable sets are required to be {\em rich} in a certain sense, then such a rich vNM stable set exists if and only if the set of efficient matchings is {\sl essentially} unique, and the set of all efficient and IR outcomes forms the unique rich vNM stable set.
\end{abstract}



\bigskip{}

\thispagestyle{empty}

\newpage{}

% \tableofcontents
% \thispagestyle{empty}

% \newpage{}

\clearpage\pagenumbering{arabic}

\section{Introduction}

In many labor markets, firms hire teams of workers, and the surplus generated by a team of workers exhibit complementarity. Such complementarities lead to settings where a stable outcome may not exist, a fundamental desiderata in matching theory. In the many-to-one job-matching model, stability was first studied by \cite{kelso1982job}, who showed the existence of a stable outcome when utility functions of firms over workers satisfy the gross substitutes condition -- see \citet{hatfield2019full,baldwin2019understanding,kojima2020job} for recent generalizations of this result.\footnote{There have been other approaches to show existence of stable matching beyond substitutable preferences of firms. \citet{che2019stable} shows that in a large market with a continuum of workers a stable matching exists even when firms' preference exhibit complementarities. \citet{rostek2020matching} establish the existence of stable outcomes in contracting environments with complementarities, allowing for multilateral agreements and externalities. In their transferable-utility setting, primitive contracts satisfy gross complementarity for every agent. These conditions do not generally hold in our many-to-one environment, where a worker can be employed by at most one firm and firms’ surplus functions are unrestricted.} 

In this paper, we seek an alternate solution: the vNM stable set, which was introduced as a set-valued solution concept in the magnum opus \citet{vnmbook}. Unlike core, where each outcome in the core is stable, individual outcomes in a vNM stable set have no meaning. It is the entire set that is stable. This is the reason \cite{vnmbook} interpret stable sets as ``standards of behavior". A vNM stable set is a set of individually rational (IR) outcomes (an outcome is a matching and payoffs to firms and workers), which are required to be consistent using a domination relation over the set of all IR outcomes.\footnote{According to this domination relation, outcome $a$ dominates outcome $b$, if there exists a firm and its matched workers in $a$ get higher utility in $a$ than in $b$ (with one agent strictly higher). This is akin to blocking the outcome $b$ in the sense of \cite{kelso1982job}.} vNM stability requires two consistency properties: (a) {\sl internal stability} -- outcomes inside the vNM stable set do not dominate each other; (b) {\sl external stability} -- each IR outcome not in the vNM stable set is dominated by some outcome inside.

In the job-matching model, a vNM stable set could be interpreted as a menu of IR outcomes/contracts (matching and non-negative payoff splits) that firms and workers may choose from. This menu does not undermine itself (internal stability) and anything not on the menu has a counter-contract on the menu. Hence, anything not in the menu cannot be included because it will be immediately countered by something already in the menu (external stability). Further, anything in the menu cannot be excluded because nothing already in the menu can counter it (internal stability). Hence, a vNM stable menu of contracts cannot exclude any of its members without losing its justification for exclusion, and cannot admit any outsider without introducing an internal conflict. 

Our main result shows that a vNM stable set in the job-matching model exists whenever the efficient matching is unique. The set of all IR and efficient outcomes forms a vNM stable set in that case. Thus, even when teams of workers exhibit significant complementarities, a vNM stable set exists, though a stable outcome may not exist. We show that if there are multiple efficient matchings, then (a) the vNM stable set may not exist; and (b) even when it exists, it may contain inefficient outcomes. However, by requiring the vNM stable set to be rich in a certain sense, we establish that the set of IR and efficient outcomes is essentially the unique vNM stable set. Hence, a rich vNM stable set mandates firms and workers to agree on a menu of contracts that contains only efficient matchings but leaves the division of surplus completely flexible.

Our richness is the following requirement. Call a vNM stable set $\mathcal{V}$ {\sl firm-rich} if for every outcome (with matching, say, $\mu$) in $\mathcal{V}$, another outcome with the same matching $\mu$ and that assigns zero payoffs to all workers belongs to $\mathcal{V}$ -- such an outcome is the firm-optimal IR outcome corresponding to $\mu$.\footnote{Given a matching $\mu$, the firm-IR optimal outcome assigns zero payoffs to all workers and gives the entire surplus from $\mu$ to firms.} We show that in a generic economy, a firm-rich vNM stable set exists if and only if there is a unique efficient matching.\footnote{The genericity requirement is a no-tie condition on the surplus generated by matching firms to teams of workers. In particular, it says that every firm generates different surplus when matched to different teams of workers.} In that case, the set of all IR and efficient outcomes forms the unique firm-rich vNM stable set. Firm-richness requires the extreme outcome, firm-IR optimal outcome, to be included in the vNM stable set. This is plausible in settings where firms have a lot of bargaining power in deciding the menu of contracts. If the economy is non-generic, a similar result holds under more richness, where the extreme outcomes giving each individual worker all the surplus is included in the menu of outcomes. For such rich vNM stable sets, we not only need the firm-IR optimal outcome but also {\sl worker-IR maximal} outcomes to be included in the vNM stable sets.

Unlike vNM stable sets, rich vNM stable sets exhibit simpler structure, and are easier to analyze. This allows us to make progress toward a characterization of rich vNM stable sets. In general, vNM stable sets are known to be difficult to analyze. Quoting \cite{aumann1985what}
\begin{quote}
    {\sl What has prevented the (vNM stable set) theory from being more successful is that it is so difficult to work with. Finding stable sets involves a new tour de force of mathematical reasoning for each game or class of games that is considered.}
\end{quote}

Our results contrast with the results in the matching literature without transfer, where a unique vNM stable set exists in the one-to-one matching model \citep{ehlers2007von,wako2010polynomial}.\footnote{Other papers studying vNM stable sets in one-to-one matching model (without transfers) include \citet{herings2017stable,dogan2026existence}.}

The job-matching model induces a transferrable utility (TU) cooperative game. The study of vNM stable sets in general cooperative games has a long history -- see \citet{aumann1985what} for a very early survey and \citet{subiza2025cooperative} for a recent contribution. This literature defines outcomes to be imputations (IR and efficient outcomes) of the cooperative game. In the job-matching model, this will loosely translate to finding a vNM stable set in the set of efficient and individually rational outcomes. This approach is taken in \cite{nunez2013von}, who analyze the assignment game (one-to-one job-matching model of \citet{shapley1971assignment}) and identify vNM stable sets. We allow for inefficient outcomes to be included in the vNM stable sets. Indeed, we show that vNM stable sets may contain inefficient outcomes. However, our partial converse result establishes that we get efficiency as a consequence of rich vNM stability.\footnote{There are other differences between our approach and the TU game approach to vNM stability -- see Remark \ref{rem:r1}.}

The domination relation used to define the vNM stable set reflects a myopic notion of blocking in the job-matching problem. An interesting direction of further research is to consider a farsighted notion of blocking. Farsighted notions of blocking and related vNM stable sets have been studied in matching literature without transfers  by~\citet{mauleon2011von,klaus2011farsighted,herings2020matching,dogan2026existence}. The nature of dominance relation used to define vNM farsighted stable set considers a sequence of blockings, and has different implications on the structure of stable sets. In the future, we plan to investigate the structure of vNM farsighted stable set of the job-matching problem.

Set-valued stability notions have been recently used in matching with incomplete information. In an important paper, \citet{Liu2014stable} introduce the notion of a self-stabilizing set and show its existence in the one-to-one matching model of \citet{shapley1971assignment} with one-sided incomplete information. Their notion is similar to the rationalizability (iterated elimination) idea in game theory but applied to their model with incomplete information. \citet{Liu2024matching} is a recent survey of this growing literature. \citet{Pourpouneh2026stable} study a similar set-valued concept of stability in the one-to-one matching model of \citet{shapley1971assignment} where agents do not observe the outcomes of the other agents. They show that a subset of efficient and IR outcomes containing the core forms the largest self-stabilizing set. While this literature and our paper look at set-valued stability solutions, there are significant differences. First, as emphasized earlier, individual outcomes have no meaning in our solution and it is only the collection that is stable. This is not the case in this literature. Second, the set-valued stability is driven by incomplete information in these papers, but we have complete information. Third, the main method to reach a stable set in these papers is an iterated elimination procedure using the uncertainty in these models. This is not the case for us.

The paper is organized as follows. Section \ref{sec:model} introduces the job-matching model. Section \ref{sec:vnm} defines stability, a domination relation among IR outcomes, and vNM stable set. Section \ref{sec:exist} proves our first result on existence of vNM stable sets. Section \ref{sec:rich} introduces rich vNM stable sets and proves our results on uniqueness of rich vNM stable sets. Appendix \ref{sec:prooftheo2} provides a detailed proof of these results, but all other omitted proofs are in Appendix \ref{sec:omapp}. We conclude in Section \ref{sec:disc} with a discussion on our results.

\section{Model}
\label{sec:model}

We study the standard model of job-matching due to \citet{kelso1982job}.
Let $F$ be the set of firms and $W$ be the set of workers. A matching is a map $\mu:W \rightarrow F \cup \{\emptyset\}$, where $\mu(w)$ denotes the firm worker $w$ is matched to -- if $\mu(w)=\emptyset$, then it denotes an unmatched worker. For every $f \in F$, we denote by $\mu(f):=\{w:\mu(w)=f\}$ the set of workers matched to firm $f$ in $\mu$.

For every $f \in F$ and every $S \subset W$, let $V_f(S)$ denote the {\bf surplus} generated by matching workers in $S$ to firm $f$. The surplus function $V_f$ may be generated from value function $v_f:2^W \rightarrow \Re_+$ of the firm (with $v_f(\emptyset)=0$) and disutility function $v_w: F \cup \{\emptyset\} \rightarrow \Re_+$ of each worker $w$:
\begin{align*}
    V_f(S) = v_f(S) - \sum_{w \in S}v_w(f)
\end{align*}
where $v_w(\emptyset)=0$.
Combined with salaries, this generates utilities for firms and workers as in \citet{kelso1982job}. However, the surplus function $V_f$ of each firm $f$ is a sufficient reduced form for our purposes. As the surplus function is generated from value of firm and disutilities of workers, (a) it can take negative values; (b) it need not be monotone.

An {\bf outcome} is a pair $(\mu,u)$, where $\mu$ is a matching and $u:W \rightarrow \Re$ corresponds to the utility assigned to {\sl workers}. An outcome specifies the utility assigned to each firm indirectly. In particular, in outcome $(\mu,u)$, each firm $f$ receives utility equal to 
$$V_f(\mu(f)) - \sum_{w \in \mu(f)}u(w)$$
Such a division of surplus is possible because utility is transferrable (using salaries) in this model. This also explicitly assumes budget-balance among a firm $f$ and its matched workers $\mu(f)$. That is, transferring surplus across firms is not possible in our model. This is consistent with much of the literature on job-matching, and makes sense because firms are independent financial entities. Also, note that by our definition of an outcome, any worker who is unmatched must get zero utility.

An economy consists of the set of firms, the set of workers, and the surplus functions:
\begin{align*}
    \Big(F,W, \{V_f\}_{f \in F} \Big)
\end{align*}

An outcome $(\mu,u)$ is {\bf individually rational (IR)} if 
\begin{align*}
    u(w) &\ge 0~\qquad~\forall~w \in W \\
    V_f(\mu(f)) - \sum_{w \in \mu(f)}u(w) &\ge 0~\qquad~\forall~f \in F,
\end{align*}
As we emphasized earlier, by the definition of an outcome, the first inequality is an equality for any worker who is unmatched.

The two constraints correspond to IR of workers and firms respectively. Let $\mathcal{I}$ be the set of all IR outcomes.

A matching $\mu^\star$ is {\bf efficient} if 
\begin{align*}
    \mu^\star \in \arg \max_{\mu} \sum_{f \in F}V_f(\mu(f))
\end{align*}
Clearly, in an efficient matching $V_f(\mu^\star(f)) \ge 0$ for all firms $f \in F$. An outcome $(\mu,u)$ is efficient if $\mu$ is efficient. Notice that the only way an outcome is {\sl inefficient} if the matching is not efficient. This is because our requirement of budget balance inside a single firm -- any outcome must split the surplus raised by a matched firm and workers between them.

\section{Stable outcomes, domination, and vNM stable sets}
\label{sec:vnm}

In this section, we define the notion of vNM stable sets. It will require us to define a relation over the set of IR outcomes. To do so, it is useful to remind the notion of blocking and stability first.

\subsection{Stable outcomes}

A firm $f \in F$ and a subset of workers $S \subseteq W$ {\bf block} an outcome $(\mu,u)$ if there exists a utility vector $\hat{u}:S \rightarrow \Re_+$ such that
\begin{align*}
    \hat{u}(w) &\ge u(w)~\qquad~\forall~w \in S \\
    V_f(S) - \sum_{w \in S}\hat{u}(w) &\ge V_f(\mu(f)) - \sum_{w \in \mu(f)}u(w)
\end{align*}
with at least one inequality strict. In other words, firm $f$ and workers in $S$ can be better off sharing the surplus $V_f(S)$ using $\hat{u}$ than the current outcome $(\mu,u)$. An outcome $(\mu,u) \in \mathcal{I}$ is {\bf stable} if it is not blocked by any $(f,S)$ pair where $f \in F$ and $S \subseteq W$ and it is individually rational. Notice that $(f,\mu(f))$ cannot block an outcome $(\mu,u)$ -- putting $S=\mu(f)$ in the blocking inequalities above and adding them gives $0 > 0$. Also, it is well known that a stable outcome may not exist, and the set of stable outcomes is also the {\sl core} of the induced transferrable utility cooperative game.

\subsection{A domination relation among IR outcomes}

The following relation among outcomes in $\mathcal{I}$ is used to define vNM stability.

\begin{defn}
    An outcome $(\mu,u) \in \mathcal{I}$ is {\bf dominated} by an outcome $(\mu',u') \in \mathcal{I}$, written as $(\mu',u') \rightarrow (\mu,u)$, if there exists a firm $f$ such that
    \begin{align*}
        u'(w) &\ge u(w)~\qquad~\forall~w \in \mu'(f) \\
        V_f(\mu'(f)) - \sum_{w \in \mu'(f)}u'(w) &\ge V_f(\mu(f)) - \sum_{w \in \mu(f)}u(w)
    \end{align*}
    with one inequality strict. We denote this as $(\mu',u') \xrightarrow[]{~f~} (\mu,u)$.
\end{defn}
Hence, $(\mu',u') \rightarrow (\mu,u)$ requires that some firm $f$ and workers $\mu'(f)$, who are matched to $f$ in $\mu'$, block $(\mu,u)$ (using $u'$). Since it is possible that two different firms evaluate these two outcomes differently, the relation $\rightarrow$ is not necessarily anti-symmetric. It also does not define a complete relation on $\mathcal{I}$. If two outcomes $(\mu,u)$ and $(\mu',u')$ are not comparable under $\rightarrow$, we write $(\mu,u) \nleftrightarrow (\mu',u')$.

For every matching $\mu$, let $\mathcal{I}_\mu$ denote the set of all IR outcomes associated with $\mu$:
\begin{align*}
    \mathcal{I}_{\mu}:=\{(\mu,u): (\mu,u) \in \mathcal{I}\}
\end{align*}
Since surplus function $V_f$ of each firm $f$ need not be non-negative, $\mathcal{I}_{\mu}$ may be empty. For instance, consider an economy with only one firm and one worker and suppose the disutility of worker is larger than the value it generates for the firm, i.e., the surplus is negative. Then the matching $\mu$ that matches the worker to the firm has no IR outcome $(\mu,u)$. Hence, $\mathcal{I}_{\mu}=\emptyset$. Here, the efficient matching is to not match the worker to the firm, in which everyone getting zero payoff is IR. In general, if $\mu^\star$ is an efficient matching, $V_f(\mu^\star(f)) \ge 0$ for all $f$ and hence, the zero payoff vector ${\bf 0}$ ($u_w=0$ for all $w$) generates the IR outcome $(\mu^\star,{\bf 0})$. Hence, $\mathcal{I}_{\mu^\star}$ is always non-empty.

\medskip

Two outcomes with the same matching are not comparable according to relation $\rightarrow$. This follows immediately because for blocking the matching needs to change. We record this as an observation.
\begin{obs}
    \label{obs:is1}
    For every matching $\mu$, $(\mu,u) \nleftrightarrow (\mu,u')$ for all $(\mu,u), (\mu,u') \in \mathcal{I}_\mu$.
\end{obs}
% \begin{proof}
%     Assume for contradiction $(\mu,u') \xrightarrow{~f~} (\mu,u)$. Then,
%     \begin{align*}
%         u'(w) &\ge u(w)~\qquad~\forall~w \in \mu(f) \\
%         V_f(\mu(f)) - \sum_{w \in \mu(f)}u'(w) &\ge V_f(\mu(f)) - \sum_{w \in \mu(f)}u(w)
%     \end{align*}
%     with one inequality strict. But adding the inequalities give us $0 > 0$, a contradiction.
% \end{proof}

\subsection{vNM stable sets}

We now define vNM stability using the relation $\rightarrow$.
\begin{defn}
    A set of outcomes $\mathcal{V} \subseteq \mathcal{I}$ is a {\bf vNM stable set} if 
    \begin{itemize}
        \item no outcome in $\mathcal{V}$ is dominated by another outcome in $\mathcal{V}$: for every $(\mu,u),(\mu',u') \in \mathcal{V}$,
        \begin{align}\label{eq:is}
            (\mu,u) \nleftrightarrow (\mu',u') \tag{\textbf{Internal stability}}
        \end{align}

        \item every outcome {\sl not in} $\mathcal{V}$ is dominated by some outcome in $\mathcal{V}$: for every $(\mu,u) \in \mathcal{I} \setminus \mathcal{V}$, there exists $(\mu',u') \in \mathcal{V}$ such that 
        \begin{align}\label{eq:exs}
            (\mu',u') \rightarrow (\mu,u) \tag{\textbf{External stability}}
        \end{align}
    \end{itemize}
\end{defn}

\begin{remark}\label{rem:r1}
    {\rm We have chosen to consider the set of IR outcomes as the set of feasible outcomes. This set will include inefficient outcomes too. This is a departure from how the TU cooperative game theory literature studies vNM stable sets. In TU games studying vNM stable sets, an outcome is an {\sl imputation} vector of players in the game, where an imputation vector is defined to be efficient (sum of imputations of players gives the value of the grand coalition of the TU game) and IR~\citep{peters2008game,nunez2013von,subiza2025cooperative}. We find no reason to restrict attention to efficient outcomes in our model. Indeed, Appendix \ref{sec:exampleapp} contains an example that has a vNM stable set with {\sl only inefficient} outcomes (where core is empty). Further, the study of vNM stable sets in the matching literature considers inefficient matchings -- see \cite{ehlers2007von}. Also, even our IR and efficient outcomes are different from imputations in TU games because we require budget-balance between every firm and its matched workers -- cross-subsidizing payoffs across firms is not allowed in our outcomes, which is consistent with \citet{kelso1982job}.} $\hfill \blacksquare{}$
\end{remark}
\medskip

If the set of stable outcomes is non-empty, it is contained in every vNM stable set. This is a standard fact but we document it as an observation.
\begin{obs}
    \label{obs:stablevnm}
    If $(\mu^\star,u^\star)$ is a stable outcome, then $(\mu^\star,u^\star) \in \mathcal{V}$ for any vNM stable set $\mathcal{V}$.
\end{obs}
\noindent The proof of this observation follows from external stability: if $(\mu^\star,u^\star) \notin \mathcal{V}$, then external stability will mean some outcome $(\mu,u) \in \mathcal{V}$ will dominate $(\mu^\star,u^\star)$; but that will mean a firm $f$ and the subset of workers $\mu(f)$ will block $(\mu^\star,u^\star)$, a contradiction to stability of $(\mu^\star,u^\star)$.

\section{Existence of vNM stable sets}
\label{sec:exist}

A vNM stable set may not exist as the following example shows.
\begin{example}[vNM stable set may not exist]
    \label{ex:examp2}
    {\rm Consider the following economy. There are three firms $F=\{f_1,f_2,f_3\}$ and three workers $W=\{w_1,w_2,w_3\}$. Firms are {\sl single-minded}: firm $f_1$ can generate positive (equal to $1$) surplus with workers $\{w_1,w_2\}$ only; firm $f_2$ can generate positive (equal to $1$) surplus with workers $\{w_2,w_3\}$ only; and firm $f_3$ can generate positive (equal to $1$) surplus with workers $\{w_3,w_1\}$ only. This is given in Table \ref{tab:t1}.\footnote{We have chosen negative surplus values for some entries in Table \ref{tab:t1}. This is only to make the analysis simpler. The example can be modified to make all surplus entries non-negative and still the vNM stable set will be empty.}
    \begin{table}[!hbt]
        \centering
        \begin{tabular}{|c||c|c|c|c|c|c|c|}
        \hline
             & $\{w_1\}$ & $\{w_2\}$ & $\{w_3\}$ & $\{w_1,w_2\}$ & $\{w_2,w_3\}$ & $\{w_3,w_1\}$ & $\{w_1,w_2,w_3\}$ \\
             \hline 
        $V_{f_1}(\cdot)$ & -1 & -1 & 0 & 1 & 0 & 0 & 0 \\
        $V_{f_2}(\cdot)$ & 0 & -1 & -1 & 0 & 1 & 0 & 0 \\
        $V_{f_3}(\cdot)$ & -1 & 0 & -1 & 0 & 0 & 1 & 0 \\
        \hline
        \end{tabular}
        \caption{Surplus of firms and workers}
        \label{tab:t1}
    \end{table}
    There are three efficient matchings: $\mu_1$ matches $f_1$ to workers $\{w_1,w_2\}$; $\mu_2$ matches $f_2$ to workers $\{w_2,w_3\}$ and $\mu_3$ matches $f_3$ to workers $\{w_3,w_1\}$. In any other matching the total surplus generated is less than or equal to zero and IR requires that all the firms and workers must get zero payoff in any outcome corresponding to such matchings. So, if $\mu \notin \{\mu_1,\mu_2,\mu_3\}$, then $(\mu,u_{w_1},u_{w_2},u_{w_3}) \in \mathcal{I}$ implies $u_{w_k}=0$ for all $k \in \{1,2,3\}$. We show that no vNM stable set exists for this example.}
\end{example}

    \begin{lemma}
        \label{lem:vnmempty}
        A vNM stable set does not exist for the economy in Example \ref{ex:examp2}.
    \end{lemma}

While vNM stable set does not exist in Example \ref{ex:examp2}, it is a {\sl non-generic} example in the sense that there are multiple efficient matchings. We now show that if a unique efficient matching exists, a vNM stable set always exists: the set of all efficient and IR outcomes is one such set.\footnote{Fixing the disutilities of workers, the set of utility functions of firms where a unique efficient matching exists has Lebesgue measure one in the set of all utility functions of firms.}

\begin{theorem}
    \label{theo:vnmstable}
    If there is a unique efficient matching, then the set of all IR and efficient outcomes is a vNM stable set.
\end{theorem}
\begin{proof}
    Internal stability of $\mathcal{I}_{\mu^\star}$ follows from Observation \ref{obs:is1}. The following lemma, whose proof is given in Appendix \ref{sec:omapp}, establishes external stability.
    \begin{lemma}
        \label{lem:order}
        Suppose $\mu$ is not an efficient matching and $(\mu,u) \in \mathcal{I}$. Then, there exists $(\mu^\star,u^\star) \in \mathcal{I}_{\mu^\star}$ such that $\mu^\star$ is efficient and $(\mu^\star,u^\star) \rightarrow (\mu,u)$.
    \end{lemma}
     Since the unique efficient matching is $\mu^\star$, any $(\mu,u) \notin \mathcal{I}_{\mu^\star}$ is inefficient. Hence, Lemma \ref{lem:order} implies there exists $(\mu^\star,u^\star) \in \mathcal{I}_{\mu^\star}$ such that $(\mu^\star,u^\star) \rightarrow (\mu,u)$.
\end{proof}

\medskip

\noindent {\sc Stable outcomes and vNM stable sets.} By Observation \ref{obs:stablevnm}, whenever the core is non-empty, it is included in every vNM stable set. We now share two observations about vNM stable sets when core is non-empty: (a) core may not be a vNM stable set; (b) a vNM stable set may contain inefficient outcomes along with core.

\begin{corollary}
    \label{cor:corenotvnm}
    Suppose $\mu^\star$ is the unique efficient matching and the core $\mathcal{C}$ is non-empty. Then, $\mathcal{C}$ is a vNM stable set if and only if $\mathcal{C}=\mathcal{I}_{\mu^\star}$. 
\end{corollary}

\begin{proof}
    The fact that $\mathcal{I}_{\mu^\star}$ is a vNM stable set follows from Theorem \ref{theo:vnmstable}. For the other direction, suppose $\mathcal{C}$ is a vNM stable set. Since every outcome in $\mathcal{C}$ is efficient, we have $\mathcal{C} \subseteq \mathcal{I}_{\mu^\star}$. Assume for contradiction $\mathcal{C} \subsetneq \mathcal{I}_{\mu^\star}$. Now, pick $(\mu^\star,u^\star) \in \mathcal{I}_{\mu^\star} \setminus \mathcal{C}$. By Observation \ref{obs:is1}, for every $(\mu^\star,u) \in \mathcal{C}$, we have $(\mu^\star,u) \nrightarrow (\mu^\star,u^\star)$. Hence, $\mathcal{C}$ violates external stability, a contradiction.
\end{proof}

Usually core is a strict subset of the set of IR outcomes corresponding to efficient matchings. Hence, when there is a unique efficient matching, core is not usually vNM stable. The following example shows that even when there are multiple efficient matchings, the core may be a strict subset of a vNM stable set and the vNM stable set may contain {\sl inefficient} matchings.

\begin{example}[vNM stable set with inefficient outcomes]
    \label{ex:vnmineff}
    {\rm Let $F=\{f\}$ and $W=\{w_1,w_2\}$. Let $$V_f(\{w_1,w_2\})=1, V_f(\{w_1\})=V_f(\{w_2\})=2.$$ Consider the four possible matchings:
    \begin{align*}
        \mu_1: \mu_1(f)=\{w_1\};~~\mu_2: \mu_2(f)=\{w_2\};~~\mu_3: \mu_3(f)=\{w_1,w_2\};~~\mu_4:\mu_4(f)=\emptyset
    \end{align*}
    Out of these, $\mu_1$ and $\mu_2$ are efficient. There are exactly two stable outcomes: $(\mu_1,0,0)$ and $(\mu_2,0,0)$. So core consists of $\mathcal{C}:=\{(\mu_1,0,0),(\mu_2,0,0)\}$. Consider the following set of inefficient outcomes.
    \begin{align*}
        \mathcal{I}_{\mu_3}^+:= \{(\mu_3,x,y): x > 0, y > 0, x+y \le 1\}
    \end{align*}
    We argue that $\mathcal{C} \cup \mathcal{I}_{\mu_3}^+$ is a vNM stable set. By Observation \ref{obs:stablevnm}, outcomes in $\mathcal{C}$ do not dominate each other. By Observation \ref{obs:is1}, outcomes in $\mathcal{I}^+_{\mu_3}$ do not dominate each other. Since firm $f$ gets payoff $2$ in outcomes in $\mathcal{C}$ and gets a maximum payoff of $1$ in outcomes in $\mathcal{I}_{\mu_3}^+$, no outcome in $\mathcal{C}$ can be dominated by outcomes in $\mathcal{I}_{\mu_3}^+$. Finally, both workers get zero payoff in outcomes in $\mathcal{C}$ but get positive payoff in outcomes in $\mathcal{I}_{\mu_3}^+$. Hence, outcomes in $\mathcal{I}_{\mu_3}^+$ cannot be dominated by outcomes in $\mathcal{C}$.

    Next, we argue external stability. Outcomes corresponding to $\mu_4$ are clearly dominated by $(\mu_1,0,0)$ (or any other outcome in $\mathcal{C} \cup \mathcal{I}_{\mu_3}^+$). Outcomes of the form $(\mu_3,x,y)$, where $xy=0$, is dominated by $(\mu_1,0,0)$ if $x=0$ and by $(\mu_2,0,0)$ if $y=0$.
    }
    $\hfill \blacksquare{}$
\end{example}

Appendix \ref{sec:exampleapp} contains another example in which the corresponding economy has an empty core but a vNM stable set containing {\sl only inefficient} outcomes exist.

\section{Toward a characterization of vNM stable sets}
\label{sec:rich}

Example \ref{ex:examp2} shows that vNM stable sets may not exist with multiple efficient matchings. The following example shows that if there are multiple efficient matchings, $\mathcal{I}_\mu$ need not be vNM stable for any efficient matching $\mu$. However, a vNM stable set exists in this example.
\begin{example}[Multiple efficient matchings]
\label{ex:examp1}
{\rm Consider the following economy. There is one firm $f$ and two workers $w$ and $w'$. The surplus function of the firm is $V_f(\{w\})=V_f(\{w'\})=1$ and $V_f(\{w,w'\})=0$. There are two efficient matchings $\mu:$ matches the firm to worker $w$ and $\mu':$ matches the firm to worker $w'$. Then, it is not difficult to see that 
\begin{align*}
    \mathcal{I}_{\mu} = \{(\mu,u(w),u(w')): u(w')=0, 0 \le u(w) \le 1 \} \\
    \mathcal{I}_{\mu'} = \{(\mu',u(w),u(w')): u(w)=0, 0 \le u(w') \le 1 \}
\end{align*}
We first argue that $\mathcal{I}_{\mu}$ is not a vNM stable set. Indeed, the outcome $(\mu',0,0)$ cannot be dominated by any outcome in $\mathcal{I}_{\mu}$. To see this, if $(\mu,x,0)$ dominates $(\mu',0,0)$, then we see that the payoff of firm is $1-x \le 1$, where $1$ is its payoff in $(\mu',0,0)$. Hence, for domination, we must have $x=0$ which gives equal payoff to firm $f$ in both outcomes. Hence, all the workers are also indifferent between the two outcomes. Hence, $\mathcal{I}_{\mu}$ is not a vNM stable set. An analogous argument shows $\mathcal{I}_{\mu'}$ is also not a vNM stable set.

Is $\mathcal{I}_\mu \cup \mathcal{I}_{\mu'}$ a vNM stable set? No. Consider $(\mu,0,0)$ and $(\mu',0,\frac{1}{2})$. Worker $w$ gets the same payoff in both outcomes and firm $f$ is strictly better in $(\mu,0,0)$. Hence, $(\mu,0,0)$ dominates $(\mu',0,\frac{1}{2})$, violating internal stability.

Is anything vNM stable? Consider the set of two outcomes $\{(\mu,0,0),(\mu',0,0)\}$. The firm and the workers are all indifferent between the outcomes. So, they are not comparable, and internally stable. For external stability, any other matching $\mu''$ has only zero payoffs for firm and two workers.\footnote{There are two other matchings, a matching that matches both the workers to the firm and a matching that matches no workers to the firm. Both these matchings generate zero surplus, and hence, by IR, must give zero payoff to the firm and workers.} Now, outcome $(\mu,0,0)$ dominates any such outcome. Also, $(\mu',0,0)$ dominates any outcome $(\mu,x,0)$ with $0 < x \le 1$. Similarly, $(\mu,0,0)$ dominates any outcome $(\mu',0,x)$ with $0 < x \le 1$. Hence, $\{(\mu,0,0),(\mu',0,0)\}$ is a vNM stable set. $\hfill \blacksquare{}$}
\end{example}

We now impose a richness on vNM stable sets. We want our vNM stable sets to be rich in the sense that if an outcome $(\mu,u)$ is part of a vNM stable set then certain other outcomes corresponding to $\mu$ should be part of the vNM stable set. One reason why we investigate rich vNM stable sets is that it allows us to establish a partial converse of Theorem \ref{theo:vnmstable}, and thus showing the salience of the rich vNM stable set identified in Theorem \ref{theo:vnmstable}. The other reason to investigate rich vNM stable sets is methodological. As we show below, {\sl rich} vNM stable sets are easier to analyze. Indeed, the sufficient condition (unique efficient matching) in Theorem \ref{theo:vnmstable} is almost necessary for the existence of rich vNM stable sets, and the rich vNM stable set identified in Theorem \ref{theo:vnmstable} is almost unique. 

We introduce two version of richness, the first one weaker than the second. The first version of richness requires that certain firm-optimal outcomes to be included in the vNM stable sets.
\begin{defn}
    A set of outcomes $\mathcal{V} \subseteq \mathcal{I}$ is {\bf firm-rich} if 
    \begin{align*}
        \Big[(\mu,u) \in \mathcal{V}\Big] \Longrightarrow \Big[ (\mu,{\bf 0}) \in \mathcal{V}\Big]
    \end{align*}
\end{defn}
Notice that the outcome $(\mu,{\bf 0})$ assigns zero payoff to all the workers and gives the entire surplus to firms. Hence, it corresponds to the {\sl firm-optimal} outcome corresponding to matching $\mu$. Firm-richness requires that if some IR outcome $(\mu,u)$ is included in $\mathcal{V}$, then the firm-optimal outcome corresponding to $\mu$ is also included in $\mathcal{V}$.

An economy is {\bf generic} if for every $f \in F$ and every $S,T \subseteq W$, we have
\begin{align*}
        \Big[V_f(S)=V_f(T)\Big] \Longrightarrow \Big[S=T\Big]
\end{align*}

Note that the set of all efficient and IR outcomes is firm-rich. Our first characterization is of firm-rich vNM stable sets in generic economies.
\begin{theorem}
    \label{theo:firmrich}
    Suppose the economy is generic. A set of outcomes $\mathcal{V}$ is a firm-rich vNM stable set if and only if 
    \begin{itemize}
        \item[(a)] there is a unique efficient matching and 
        \item[(b)] $\mathcal{V}$ is the set of all efficient and IR outcomes.
    \end{itemize}
\end{theorem}

Theorem \ref{theo:vnmstable} already proves one direction of Theorem \ref{theo:firmrich}. Hence, what remains to be proved is that existence of a firm-rich vNM stable set implies the economy must have a unique efficient matching, and that firm-rich vNM stable set is precisely the one identified in Theorem \ref{theo:vnmstable}. We give a combined proof of Theorem \ref{theo:firmrich} and \ref{theo:multieff} in Appendix \ref{sec:prooftheo2}. To see how generic economy with firm-rich vNM stable set helps in the proof, let $\mathcal{V}$ be a firm-rich vNM stable set. Suppose $\mathcal{V}$ contains outcomes $(\mu,u)$ and $(\mu',u')$. Firm-richness implies $(\mu,{\bf 0}),(\mu',{\bf 0}) \in \mathcal{V}$. Now, pick any firm $f$ and suppose $\mu(f) \ne \mu'(f)$. If $V_f(\mu(f)) > V_f(\mu'(f))$, then $(\mu,{\bf 0}) \xrightarrow[]{~f~}(\mu',{\bf 0})$, which will contradict internal stability. Similarly, if $V_f(\mu(f)) < V_f(\mu'(f))$, then $(\mu',{\bf 0}) \xrightarrow[]{~f~}(\mu,{\bf 0})$. Hence, $V_f(\mu(f))=V_f(\mu'(f))$ for all $f$. Generic economy assumption allows us to then conclude $\mu(f)=\mu'(f)$ for all $f$. Hence, we can conclude that in any firm-rich vNM stable set in a generic economy, outcomes corresponding to a single matching $\mu$ must exist. By Observation \ref{obs:is1}, such a vNM stable set must be $\mathcal{I}_{\mu}$ -- the set of all IR outcomes corresponding to the matching $\mu$. The fact that $\mu$ is the unique efficient matching is significantly more complicated to show. The proof of this part relies on an elegant result in network flows by David Gale, for which we interpret some of our external stability conditions existence of flows in an appropriate network.

\begin{remark}\label{rem:weakrich}
{\rm Firm-richness requires inclusion of extreme outcomes in the vNM stable set. As a result, we show the almost necessity of unique efficient outcomes in vNM stable sets and an almost unique characterization of the vNM stable set identified in Theorem \ref{theo:vnmstable}. However, it is arguable that such extreme outcomes should {\sl not} be included in a vNM stable set, given that it is supposed to reflect ``standard social behavior" (in the language of \citet{vnmbook}). While it is difficult to characterize all vNM stable sets, we believe that firm-richness is plausible. In economies where firms have high bargaining power, firm-optimal outcomes may have to be included in any menu of contracts.} \hfill $\blacksquare{}$
\end{remark}

We now relax the generic economy assumption. By requiring more richness in the vNM stable set, we derive a characterization analogous to Theorem \ref{theo:firmrich}.

 We say a matching $\mu$ is {\bf IR-feasible} if $\mathcal{I}_\mu$ is non-empty. For any set of outcomes $\mathcal{V} \subset \mathcal{I}$, denote by $\mathcal{M}(\mathcal{V})$ the set of matchings involved in $\mathcal{V}$:
\begin{align*}
    \mathcal{M}(\mathcal{V}):= \{\mu: (\mu,u) \in \mathcal{V}\}
\end{align*}

Without genericity, a firm-rich vNM stable set $\mathcal{V}$ may contain outcomes from more than one matching. To characterize such sets, we strengthen our richness requirement. We now consider worker-IR-maximal outcomes, and the consequence of including them in vNM stable sets. For every worker $w$, let 
\begin{align*}
    u(w) := 
    \begin{cases}
        V_{\mu(w)}(\mu(\mu(w))) & \textrm{if}~\mu(w) \ne \emptyset \\
        0 & \textrm{otherwise}
    \end{cases}
\end{align*}
and $u(w'):=0$ for all $w' \ne w$. Clearly, $(\mu,u) \in \mathcal{I}_{\mu}$. We refer to $(\mu,u)$ as {\bf $w$-IR-maximal} outcome for $\mu$.\footnote{Notice that this outcome not only gives the maximum possible payoff to $w$ but also gives zero payoff to all other workers. In that sense, it is choosing a $w$-IR-maximal outcome that is worst for other workers.}

\begin{defn}
    A set of outcomes $\mathcal{V} \subset \mathcal{I}$ is {\bf rich} if it is firm-rich and for every $\mu \in \mathcal{M}(\mathcal{V})$ and for every worker $w$, the $w$-IR-maximal outcome for $\mu$ belongs to $\mathcal{V}$.
\end{defn}

As we will show below, without genericity, a rich vNM stable set $\mathcal{V}$ may contain outcomes corresponding to more than one matching, but these matchings are essentially unique in the following sense.
\begin{defn}
    A set of IR-feasible matchings $\mathcal{M}'$ is {\bf essentially unique} if for all $\mu,\mu'\in\mathcal{M}'$ and for all  $f\in F$, 
    \begin{align*}
        \Big[ \mu(f) \ne \mu'(f)\Big] \Longrightarrow \Big[V_f(\mu(f)) = V_f(\mu'(f)) = 0\Big]
    \end{align*}
\end{defn}
For any matching $\mu$, let $V(\mu):=\sum_{f \in F}V_f(\mu(f))$ denote the total surplus generated from $\mu$. Note that if $\mathcal{M'}$ is essentially unique, then for all $\mu,\mu' \in \mathcal{M'}$, we have $V(\mu)=V(\mu')$.

\begin{theorem}
\label{theo:multieff}
A set of outcomes $\mathcal{V}$ is a rich vNM stable set if and only if 
\begin{itemize}
\item[(a)] $\mathcal{M}(\mathcal{V})$ is essentially unique and 
\item[(b)] $\mathcal{V}$ is the set of all efficient and IR outcomes.
\end{itemize}
\end{theorem}

Thus, if we move beyond generic economies, we can establish the analogue of Theorem \ref{theo:firmrich} under additional richness. The following is an immediate corollary of Theorem \ref{theo:multieff}.
\begin{corollary}
    \label{cor:effuni}
    Suppose $\mu$ is an IR-feasible matching. Then, $\mathcal{V} \subseteq \mathcal{I}_\mu$ is vNM stable if and only if $\mu$ is the unique efficient matching and $\mathcal{V}=\mathcal{I}_\mu$.
\end{corollary}
The proof of necessity direction of Corollary \ref{cor:effuni} follows because Observation \ref{obs:is1} and external stability imply $\mathcal{V}=\mathcal{I}_{\mu}$, and hence, $\mathcal{V}$ is a rich vNM stable set. Theorem \ref{theo:multieff} then implies the result.

\section{Conclusion}
\label{sec:disc}

Our main motivation for studying vNM stable sets was that stable outcomes are not guaranteed to exist in this model, specially in a large class of economically meaningful settings where teams of workers exhibit strong complementarities. By extending this to a collective stability notion, we have {\sl almost} resolved the existence issue. Our results suggest that in this model, if firms and workers were to choose a collection of outcomes (contracts) using the principles of vNM stability, they are {\sl almost} forced to choose only efficient outcomes but have complete flexibility in the division of surplus.

% \section{Concluding remarks}
% \label{sec:future}

\newpage

{\small{}

}

% {\small{}  \bibliographystyle{ecta}
% \bibliography{vnm_references}
%  }

\newpage
\appendix

\section{Proofs of Theorems \ref{theo:firmrich} and \ref{theo:multieff}}
\label{sec:prooftheo2}

We give a proof of Theorems \ref{theo:firmrich} and \ref{theo:multieff} here. The main ingredient of the proofs is the following proposition about the domination relation, which does not assume any kind of richness, and is of indepenendent interest.
\begin{proposition}
    \label{prop:domeff}
    Suppose $\mu^\star$ is an efficient matching and $\mu$ is an IR-feasible matching. Then, there exists $(\mu^\star,u^\star) \in \mathcal{I}_{\mu^\star}$ such that 
    \begin{align*}
        (\mu,u) \nrightarrow (\mu^\star,u^\star)~\qquad~\forall~(\mu,u) \in \mathcal{I}_{\mu}
    \end{align*}
\end{proposition}

The proof of Proposition \ref{prop:domeff}, which is given below after some preparations for the proof is done, hinges on a characterization of the conditions under which an outcome $(\mu',u') \in \mathcal{I}$ is not dominated by any outcome $(\mu,u) \in \mathcal{I}_\mu$, where $\mu \ne \mu'$. This is done in Lemma \ref{lem:domchar}, whose proof is in Appendix \ref{sec:omapp}.
    \begin{lemma}\label{lem:domchar}
       Let $\mu'$ and $\mu$ be two distinct IR-feasible matchings. Suppose  $(\mu',u')\in \mathcal{I}_{\mu'}$. Then, the following are equivalent:
       \begin{enumerate}
           \item $(\mu,u) \nrightarrow (\mu',u')$ for all $(\mu,u) \in \mathcal{I}_{\mu}$

           \item For all $f \in F$,
            \begin{align}
                V_f(\mu(f))-\sum_{w\in \mu(f)}u'(w) \le V_f(\mu'(f))-\sum_{w\in \mu'(f)}u'(w) \label{eq:dcc1}
            \end{align}
       \end{enumerate}
    \end{lemma}

    Fix any matching $\mu'$. Lemma \ref{lem:domchar} gives a clean linear system to verify when there exists $u'$ such that $(\mu',u') \in \mathcal{I}_{\mu'}$ and  $(\mu,u) \nrightarrow (\mu',u')$ for every $(\mu,u) \in \mathcal{I}_{\mu}$. We rewrite this linear system below. 
    
    To do so, we introduce a dummy firm $f_0$ such that $V_{f_0}(S)=0$ for all $S \subset W$. We assume that any worker who is unmatched in a matching is matched to $f_0$. Without loss of generality, we assume that $f_0 \in F$ and each worker is matched to some firm in $F$ (possibly to $f_0$) in every matching, where $\mu(w)=f_0$ for a matching $\mu$ would indicate that worker $w$ is unmatched. Now, consider the following system of inequalities.
      \begin{align}
       \sum_{w\in \mu'(f)}u'(w) -\sum_{w\in \mu(f)}u'(w) &\le V_f(\mu'(f))-V_f(\mu(f))~\qquad~\forall~f \in F \label{eq:es} \tag{\textbf{ES}} \\
       \sum_{w \in \mu'(f)}u'(w) &\le V_f(\mu'(f))~\qquad~\forall~f \in F \label{eq:ir} \tag{\textbf{IR}} \\
       u'(w) &\ge 0~\qquad~\forall~w \in W \nonumber
   \end{align}
   Constraints (\ref{eq:ir}) ensures that a feasible $u'$ satisfies $(\mu',u') \in \mathcal{I}$. Constraints (\ref{eq:es}) is the same as the constraints in Lemma \ref{lem:domchar}. Note that if a worker $w$ is unmatched in $\mu'$ (that is matched to $f_0$), then $u'(w) \ge 0$ and (\ref{eq:ir}) along with the fact that $V_{f_0}(\mu'(f_0))=0$ implies that $u'(w)=0$.

   If $u'$ is a feasible solution to (\ref{eq:es}) and (\ref{eq:ir}), then there exists $(\mu',u') \in \mathcal{I}_{\mu'}$ such that $(\mu,u) \nrightarrow (\mu',u')$ for all $(\mu,u) \in \mathcal{I}_\mu$. Conversely, if there exists $(\mu',u') \in \mathcal{I}_{\mu'}$, it is already a feasible solution to (\ref{eq:ir}) and if it is such that $(\mu,u) \nrightarrow (\mu',u')$ for all $(\mu,u) \in \mathcal{I}_\mu$, then by Lemma \ref{lem:domchar}, it is a feasible solution to (\ref{eq:es}).

    To verify feasibility of (\ref{eq:es}) and (\ref{eq:ir}) we represent this as a {\sl demand graph}, whose flows will represent a feasible solution. Our demand graph $\mathcal{G}$ has the following components: $\mathcal{G} \equiv (V,E,c,b)$ is a {\bf demand graph}, where
    \begin{itemize}
        \item $V$ is a finite set of vertices
        \item $E \subseteq \{(x,y):x, y \in V\}$ is a finite set of directed edges
        \item $c:E \rightarrow \Re_+$ specifies a non-negative capacity for each edge in $E$
        \item $b:V \rightarrow \Re$ specifies {\sl divergence} at each vertex -- if $b(x) > 0$, then $x$ is a {\sl demand} vertex and if $b(x) < 0$, then $x$ is a {\sl supply} vertex
    \end{itemize}

    A {\sl flow} in $\mathcal{G}$ is a map $\phi:E \rightarrow \Re_+$ satisfying the following feasibility constraints:
    \begin{align*}
        \phi(x,y) &\le c(x,y)~\qquad~\forall~(x,y) \in E \\
        \underbrace{\sum_{(y,x) \in E}\phi(y,x)}_{\textrm{In flow}} - \underbrace{\sum_{(x,y) \in E}\phi(x,y)}_{\textrm{Out flow}} &= b(x)~\qquad~\forall~x \in V 
    \end{align*}

    We fix matchings $\mu$ and $\mu'$ and construct the specific demand graph $\mathcal{G}$ as follows:
    \begin{itemize}
        \item {\bf Vertices.} There is a vertex for each worker $w \in W$. Denote such a vertex as $w$ itself for each $w \in W$. There is a pair of vertices for each firm $f \in F$. Denote such vertices as $f,\hat{f}$ for each firm $f \in F$. The set of vertices $\hat{f}$ for each $f$ will be referred to as the auxiliary vertices corresponding to firms. We will denote the set of all auxiliary vertices as $\widehat{F}$. There is another vertex $v_0$ for handling slacks in constraints (\ref{eq:es}). So, in total there are $|W|+2|F|+1$ vertices.

        \item {\bf Edges.} For every $w \in W$, there is an edge from $w$ to $f$ if $\mu'(w)=f$. There is an edge from $f$ to $\hat{f}$ for each $f \in F$. For each $f \in F$, there is an edge from $\hat{f}$ to $w$ if $\mu(w)=f$. There is an edge from $v_0$ to $\hat{f}$ for every $f \in F$. Note that each vertex $w$ has exactly one outgoing edge (to $\mu'(w)$) and exactly one incoming edge (from $\mu(w)$).\footnote{This is because each worker is matched to some firm, possible $f_0$.} 

        \item {\bf Capacities.} All edges have very high capacity (to be precise more than $V(\mu')$) except edges $(f,\hat{f})$ for each $f \in F$:
        \begin{align*}
            c(f,\hat{f}):=V_f(\mu'(f))
        \end{align*}

        \item {\bf Divergences.} Divergence at vertex $w \in W$ is zero: $b(w)=0$. Divergence at the two vertices corresponding to any firm $f \in F$ are: $b(f)=0$ but
        \begin{align*}
            b(\hat{f}):= V_f(\mu'(f)) - V_f(\mu(f))
        \end{align*}
        Divergence at $v_0$ is 
        \begin{align*}
            b(v_0) := V(\mu)-V(\mu')
        \end{align*}
        where for any matching $\hat{\mu}$, $V(\hat{\mu}) \equiv \sum_{f\in F} V_f(\hat{\mu}(f))$.
    
        Note that $\sum_{x \in V}b(x)=0$.
    \end{itemize}

\begin{figure}[!hbt]
        \centering
        \includegraphics[width=0.5\linewidth]{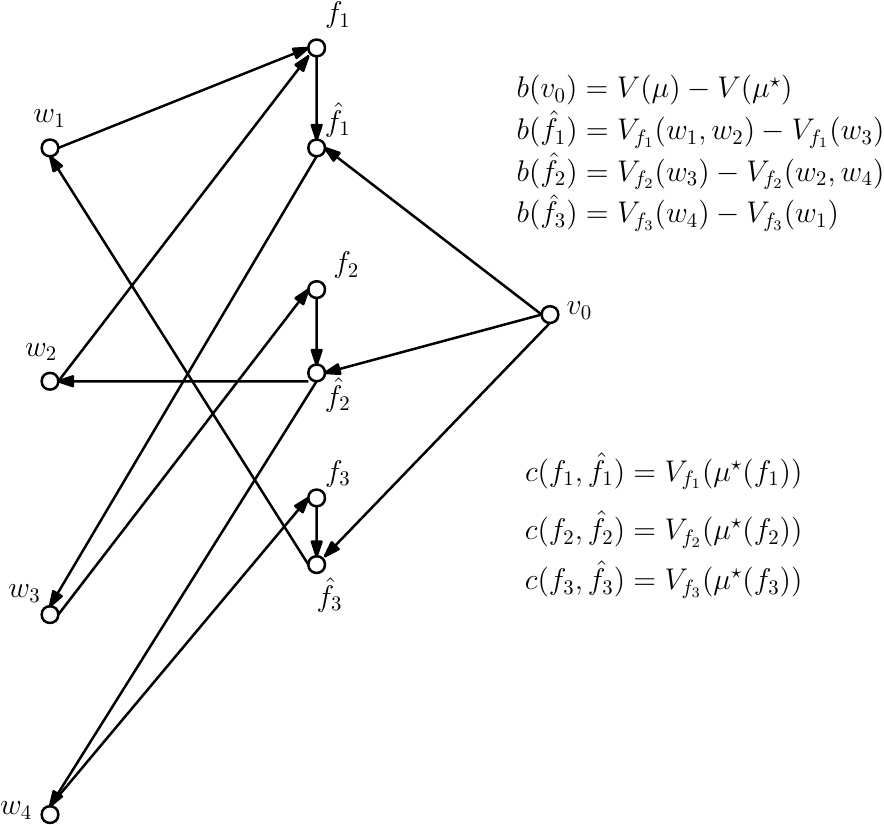}
        \caption{Demand network}
        \label{fig:gale}
    \end{figure}

    As an example consider $W=\{w_1,w_2,w_3,w_4\}$ and $F=\{f_1,f_2,f_3\}$. Suppose $\mu$ and $\mu'$ are as follows:
    \begin{align*}
        \mu'(w_1)=f_1,\mu'(w_2)=f_1,\mu'(w_3)=f_2,\mu'(w_4)=f_3 \\
        \mu(w_1)=f_3,\mu(w_2)=f_2,\mu(w_3)=f_1,\mu(w_4)=f_2
    \end{align*}
    The corresponding demand graph is shown in Figure \ref{fig:gale}
    
    The following lemma makes the connection between demand graph and the system of inequalities in (\ref{eq:es}) and (\ref{eq:ir}) precise.
    \begin{lemma}
        \label{lem:flow}
        Suppose $\mu$ and $\mu'$ are two IR-feasible matchings. A feasible solution to (\ref{eq:es}) and (\ref{eq:ir}) exists if and only if a flow in the demand graph $\mathcal{G}$ exists.
    \end{lemma}
    
    We use Lemma \ref{lem:flow}, which is proved in Appendix \ref{sec:omapp}, to show sufficiency of efficient matching for (\ref{eq:es}) and (\ref{eq:ir}) to be feasible. The proof of Proposition \ref{prop:domeff} can now be completed using a celebrated theorem of David Gale on existence of flows in a demand graph.\footnote{David Gale's existence theorem has been used to prove an important extension of Border's reduced form auction result \citep{border1991implementation} -- see \citet{che2013generalized}.}

\medskip

    \noindent {\bf Proof of Proposition \ref{prop:domeff}.} Let $\mu$ and $\mu^\star \equiv \mu'$ be two IR-feasible matchings (to be consistent with the notation in the earlier lemmas, we will denote $\mu' \equiv \mu^\star$). We show that if $\mu'$ is an efficient matching, then a flow in directed graph $\mathcal{G}$ (corresponding to $\mu$ and $\mu^\star$) exists. Lemma \ref{lem:flow} will then imply that a feasible solution to (\ref{eq:es}) and (\ref{eq:ir}) exists. Denote this solution as $(\mu^\star,u^\star)$. and Lemma \ref{lem:domchar} will imply that for all $(\mu,u) \in \mathcal{I}_\mu$, we have $(\mu,u) \nrightarrow (\mu^\star,u^\star)$. This will complete the proof.
    
    To show that a flow in directed graph $\mathcal{G}$ exists, we invoke a seminal result due to David Gale for this. We introduce some notations first. A {\sl cut} in $\mathcal{G}$ is a partitioning of the set of vertices into two parts: $(S,V \setminus S)$. Without loss of generality, we denote a cut by $S \subseteq V$. The capacity of a cut is the sum of capacities of all incoming edges to the cut:
    \begin{align*}
        c(S) := \sum_{(x,y) \in E:x \notin S,y \in S}c(x,y)~\qquad~\forall~S \subseteq V
    \end{align*}
    The divergence of a cut is the sum of divergences of vertices in the cut:
    \begin{align*}
        b(S) := \sum_{x \in S}b(x)~\qquad~\forall~S \subseteq V
    \end{align*}

    \begin{theorem}[David Gale's Demand Theorem]
        Suppose $\mathcal{G}$ is a directed graph such that $b(V)=0$. Then, a flow exists in $\mathcal{G}$ if and only if for every cut $S \subseteq V$, we have 
        \begin{align*}
            b(S) &\le c(S)
        \end{align*}
    \end{theorem}
    By David Gale's theorem, it is enough to show that for any cut $S \subseteq V$ in $\mathcal{G}$, we have $b(S) \le c(S)$. We will consider three kinds of cuts. \\

    \noindent {\sc Case 1.} This is the case where there is some incoming edge to $S$ with high capacity. Hence, $c(S)$ is a very high number and $c(S) \ge b(S)$. \\

    \noindent {\sc Case 2.} There is no high capacity  edge incoming to $S$ and $v_0 \notin S$. In that case, each auxiliary vertex $\hat{f} \notin S$ -- else there would be a high capacity edge $(v_0,\hat{f})$ into $S$. Since divergence is non-zero only for vertices in $\widehat{F} \cup \{v_0\}$, we conclude that $b(S)=0$. Since capacities are non-negative, $c(S) \ge b(S)$. \\

    \noindent {\sc Case 3.} There is no high capacity  edge incoming to $S$ and $v_0 \in S$. We are going to partition auxiliary (firm)-vertices in $S$ into two parts: $(\widehat{F}' \cup \widehat{F}'') \subset S$ with $\widehat{F}' \cap \widehat{F}'' = \emptyset$. For every $\hat{f} \in \widehat{F}'$, the vertex $f \in S$, and hence, edge $(f,\hat{f})$ is not incoming to $S$. For every $\hat{f} \in \widehat{F}''$, vertex $f \notin S$, and hence, edge $(f,\hat{f})$ is incoming to $S$.\footnote{This case also covers the case where there are no incoming edges to $S$, and hence, $c(S)=0$. In that case $F'' = \emptyset$.} That is $\widehat{F}'' \subset S$ and $F'' \subset V \setminus S$. Partition the set of workers also:
    \begin{align*}
        W':= \bigcup_{f' \in F'}\mu'(f'),~~~~~W'':= \bigcup_{f'' \in F''}\mu'(f''),~~~~~~W \setminus (W' \cup W'')
    \end{align*}
    Since $S$ has no incoming-edge with high capacity, two things must hold true:
    \begin{itemize}
        \item $\mu'(w) \in F \setminus (F' \cup F'')$ for all $w \in W \setminus (W' \cup W'')$,

        \item $W' \subset S$ and for every $w \in W'$, $\mu(w) \in F' \cup F''$.

    \end{itemize}

\begin{figure}[!hbt]
        \centering
        \includegraphics[width=0.5\linewidth]{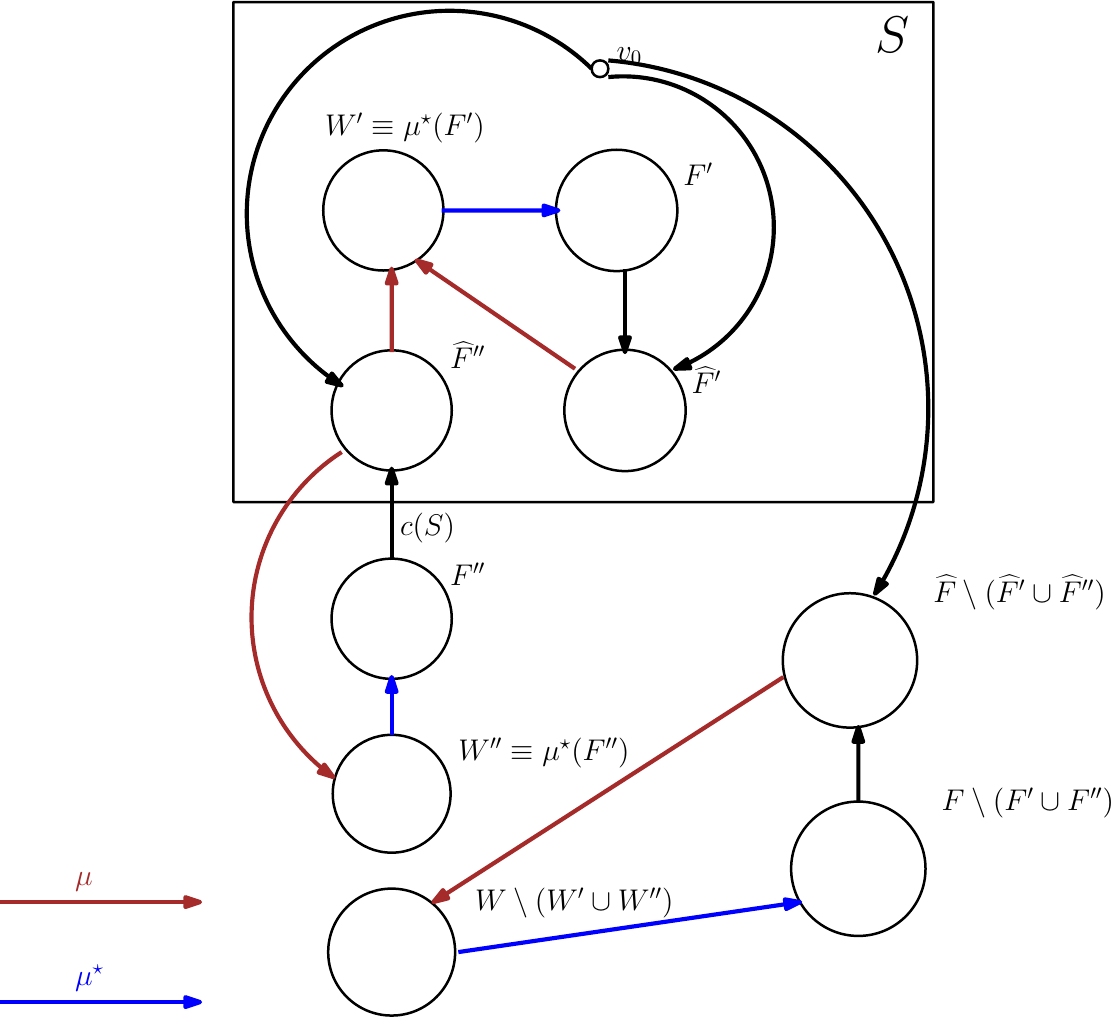}
        \caption{Representation of cut $S$}
        \label{fig:cut}
    \end{figure}

    An illustration of cut $S$ is shown in Figure \ref{fig:cut}.

The divergence of $S$ equals
     \begin{align}
         b(S) &= b(v_0) + \sum_{\hat{f} \in \widehat{F}'}b(\hat{f}) + \sum_{\hat{f} \in \widehat{F}''}b(\hat{f}) \nonumber \\
         &= b(v_0) + \sum_{f \in F'}\big[V_f(\mu'(f)) - V_f(\mu(f))\big] + \sum_{f \in F''}\big[V_f(\mu'(f)) - V_f(\mu(f))\big] \nonumber \\
          &= \big[V(\mu) - V(\mu')\big] + \sum_{f \in F'}\big[V_f(\mu'(f)) - V_f(\mu(f))\big] + \sum_{f \in F''}\big[V_f(\mu'(f)) - V_f(\mu(f))\big] \nonumber\\
          &=\sum_{f \in F\setminus (F'\cup F'')}\big[V_f(\mu(f)) - V_f(\mu'(f))\big].\nonumber
     \end{align}
     The capacity of $S$ equals $c(S)=\sum_{f\in F''} c(f,\hat{f})=\sum_{f\in F''} V_f(\mu'(f)) $ as the only incoming edges to $S$ are of the form $(f,\hat{f})$ where $f \in F''$.
   %    Consider a candidate matching $\hat{\mu}$ as follows:
   % \begin{align*}
   %     \hat{\mu}(f) = 
   %     \begin{cases}
   %         \mu'(f) &~\textrm{if}~f \in F' \\
   %         \emptyset &~\textrm{if}~f \in F''\\
   %         \mu(f) &~\textrm{if}~f \in F\setminus (F'\cup F'')
   %     \end{cases}
   % \end{align*}

   Consider a candidate matching $\hat{\mu}$ defined as follows. For every $f\in F\setminus\{f_0\}$, let
   \begin{align*}
    \hat{\mu}(f)=
    \begin{cases}
    \mu'(f) & \text{if } f\in F',\\
    \emptyset & \text{if } f\in F'',\\
    \mu(f) & \text{otherwise.}
    \end{cases}
    \end{align*}
    All remaining workers are matched to the dummy firm $f_0$, that is, $\hat{\mu}(f_0) =W\setminus\big(\bigcup_{f\in F\setminus\{f_0\}}\hat{\mu}(f)\big).$
    
   Since for every $w \in W'$, $\mu(w) \in F' \cup F''$, the matching $\hat{\mu}$ is well-defined. As $\mu'$ is efficient,
   \begin{align*}
       0 \ge V(\hat{\mu}) - V(\mu') = \sum_{f \in F\setminus (F'\cup F'')}\big[V_f(\mu(f)) - V_f(\mu'(f))\big] -\sum_{f\in F''}V_f(\mu'(f))=b(S)-c(S).
   \end{align*}
   Hence, it follows that $b(S)\leq c(S)$. 
   $\hfill \blacksquare{}$

\medskip

We now have all the tools to prove Theorems \ref{theo:firmrich} and \ref{theo:multieff}.

\medskip

\noindent {\bf Proof of Theorems \ref{theo:firmrich} and \ref{theo:multieff}.} \\

\noindent \textbf{Sufficiency}: The sufficiency direction of Theorem \ref{theo:multieff} implies the sufficiency direction of Theorem \ref{theo:firmrich}. Hence, we only give a proof of sufficiency direction of Theorem \ref{theo:multieff}. Let $\mathcal{M}^{\mathrm{eff}}$ be the set of all efficient matchings. Suppose $\mathcal{M}^{\rm eff}$ is essentially unique, and let 
\begin{align*}
    \mathcal{V} := \bigcup_{\mu\in\mathcal{M}^{\mathrm{eff}}}\mathcal{I}_{\mu}
\end{align*}
We show $\mathcal{V}$ is a vNM stable set in two steps. \\

\noindent {\sc Internal stability.} Consider any two efficient matchings $\mu,\mu' \in \mathcal{M}^{\rm eff}$. Consider any $f \in F$. Since $\mathcal{M}^{\rm eff}$ is essentially unique, $V_f(\mu(f)) = V_f(\mu'(f))$. If $\mu(f)=\mu'(f)$, then for every $(\mu',u') \in \mathcal{I}_{\mu'}$, we have
\begin{align}\label{eq:is11}
    V_f(\mu(f)) - \sum_{w \in \mu(f)}u'(w) &= V_f(\mu'(f)) - \sum_{w \in \mu'(f)}u'(w)
\end{align}
Lemma \ref{lem:domchar} implies $(\mu,u) \nrightarrow (\mu',u')$ for all $(\mu,u) \in \mathcal{I}_\mu$. If $\mu(f) \ne \mu'(f)$, then $V_f(\mu(f)) = V_f(\mu'(f)) = 0$ (by essential uniqueness). Hence, for every $(\mu',u') \in \mathcal{I}_{\mu'}$, we have $u'(w)=0$ for all $w \in \mu'(f)$ and using the fact that $u'(w) \ge 0$ for all $w$, equation (\ref{eq:is11}) holds with inequality (LHS $\le$ RHS). Hence, Lemma \ref{lem:domchar} implies $(\mu,u) \nrightarrow (\mu',u')$ for all $(\mu,u) \in \mathcal{I}_\mu$. Finally, if $\mu=\mu'$, Observation \ref{obs:is1} implies $(\mu,u) \nrightarrow (\mu',u')$ for all $(\mu,u) \in \mathcal{I}_\mu$. This shows that $\mathcal{V}$ is internally stable. \\

\noindent {\sc External stability.} Consider any outcome $(\mu,u) \in \mathcal{I} \setminus \mathcal{V}$. Hence, $\mu \notin \mathcal{M}^{\rm eff}$. By Lemma \ref{lem:order}, there exists an efficient matching $\mu^\star\in\mathcal{M}^{\mathrm{eff}}$ and an outcome $(\mu^\star,u^\star)\in\mathcal{I}_{\mu^\star}$ such that $(\mu^\star,u^\star)\rightarrow(\mu,u)$. Hence, external stability follows. \\

\noindent \textbf{Necessity}: We prove the necessity directions of Theorems \ref{theo:firmrich} and \ref{theo:multieff} together. Suppose $\mathcal{V}$ is a vNM-stable set. The following claim clarifies the role firm-richness in a vNM stable set.

\begin{claim}
    \label{cl:frich}
    Suppose $\mathcal{V}$ is a firm-rich vNM-stable set. Then, for every $\mu,\mu' \in \mathcal{M}(\mathcal{V})$, we have
    \begin{align}
        V_f(\mu(f))=V_f(\mu'(f))~\qquad~\forall~f \in F \label{eq:esuni1}
    \end{align}
    Further, if the economy is generic $\mu=\mu'$ (i.e., $\mathcal{M}(\mathcal{V})$ has a unique matching).
\end{claim}
\begin{proof}
    Suppose $\mathcal{V}$ is a firm-rich vNM-stable set. Consider two distinct matchings $\mu,\mu'\in\mathcal M(\mathcal{V})$. Since $\mathcal{I}_{\mu}$ and $\mathcal{I}_{\mu'}$ are non-empty, individual rationality implies $V_f(\mu(f))\geq 0$ and $V_f(\mu'(f))\geq 0$ for every $f\in F$. Consider $u$ and $u'$ such that $u(w)=u'(w)=0$ for every $w\in W$. Note that $u$ is firm-IR-optimal for $\mu$ and $u'$ is firm-IR-optimal for $\mu'$. Hence, we have $(\mu,u),(\mu',u') \in \mathcal{V}$. If for some $f \in F$, $V_f(\mu(f)) > V_f(\mu'(f))$, then $(\mu,u) \xrightarrow[]{~f~}(\mu',u')$, contradicting internal stability. Hence, $V_f(\mu(f))\le V_f(\mu'(f))$ for all $f \in F$. By interchanging the roles of $(\mu,u)$ and $(\mu',u')$, we obtain $V_f(\mu'(f))\le V_f(\mu(f))$ for all $f \in F$. Therefore, (\ref{eq:esuni1}) holds.

    If the economy is generic, by definition for every $f$, we have $\mu(f)=\mu'(f)$, and hence, $\mu=\mu'$.
\end{proof}

By Claim \ref{cl:frich}, if $\mathcal{V}$ is firm-rich, then for every $\mu,\mu' \in \mathcal{M}(\mathcal{V})$, we have
\begin{align}
    V_f(\mu(f))=V_f(\mu'(f))~\qquad~\forall~f \in F \label{eq:esuni11}
\end{align}
Further, if the economy is generic, then $\mathcal{M}(\mathcal{V})$ contains a unique matching.

If the economy is not generic, then assume $\mathcal{V}$ is rich. Suppose now that for some $f\in F$ we have $V_f(\mu(f))=V_f(\mu'(f))=p>0 $ but $\mu(f)\neq\mu'(f)$.\footnote{If our economy is without ties as in Remark \ref{rem:weakrich}, we do not need this part as $\mathcal{M}(\mathcal{V})$ will have only one matching, and we can establish essential uniqueness with weaker richness requirement: firm-IR-optimal outcomes are required to be included in $\mathcal{V}$.} Without loss of generality, suppose there is some $\tilde{w}\in\mu(f)\setminus\mu'(f)$. Let $\tilde{u}$ be a utility vector such that $\tilde{u}(\tilde{w})=p, \tilde{u}(w)=0$ for all $w \ne \tilde{w}$. Clearly, $\tilde{u}$ is such that $(\mu,\tilde{u})\in\mathcal I_\mu$. Further, $(\mu,\tilde{u})$ is $\tilde{w}$-IR-maximal for $\mu$. Hence, by richness, $(\mu,\tilde{u}) \in \mathcal{V}$. Moreover, all workers in $\mu'(f)$ receive zero payoff in $(\mu',u')$, while firm $f$ receives payoff $p$ at $(\mu',u')$ and payoff zero at $(\mu,\tilde{u})$. Hence, $(\mu',u')\xrightarrow[]{~f~}(\mu,\tilde{u})$, contradicting internal stability (since both $(\mu',u'), (\mu,\tilde{u}) \in \mathcal{V}$). Thus, $V_f(\mu(f))=V_f(\mu'(f))>0 $ implies that $\mu(f)=\mu'(f)$. Along with (\ref{eq:esuni11}), this proves that $\mathcal{M}(\mathcal{V})$ is essentially unique.

We show that $\mathcal{M}(\mathcal{V})=\mathcal{M}^{\mathrm{eff}}$ (in case of generic economy, this will show that $\mathcal{M}(\mathcal{V})$ contains a unique efficient matching, as is necessary of Theorem \ref{theo:firmrich}). First, note that by adding (\ref{eq:esuni11}) for all $f \in F$, we see that
\begin{align}\label{eq:surplus}
    V(\mu) = V(\mu')~\qquad~\forall~\mu,\mu' \in \mathcal{M}(\mathcal{V})
\end{align}
In other words, all matchings in $\mathcal{M}(\mathcal{V})$ generate the same surplus.

Suppose that there exists $\mu^{\star} \in \mathcal{M}^{\mathrm{eff}} \setminus \mathcal{M}(\mathcal{V})$. Consider $\bar{\mu}\in\mathcal{M}(\mathcal{V})$. By Proposition \ref{prop:domeff}, there exists $(\mu^{\star},u^{\star})\in\mathcal{I}_{\mu^{\star}}$ such that $(\bar{\mu},\bar{u}) \nrightarrow (\mu^\star,u^\star)$ for all $(\bar{\mu},\bar{u}) \in \mathcal{I}_{\bar{\mu}}$. 

We now show that this implication spreads to all $\mu \in \mathcal{M}(\mathcal{V})$ since $\mathcal{M}(\mathcal{V})$ is essentially unique. To see this, consider any $\mu\in\mathcal{M}(\mathcal{V})$. Since $(\bar{\mu},\bar{u}) \nrightarrow (\mu^\star,u^\star)$ for all $(\bar{\mu},\bar{u}) \in \mathcal{I}_{\bar{\mu}}$, by Lemma \ref{lem:domchar}, for all $f\in F$, 
\begin{align*}
    V_f(\bar{\mu}(f))-\sum_{w\in\bar{\mu}(f)}u^{\star}(w) \leq V_f(\mu^{\star}(f))-\sum_{w\in\mu^{\star}(f)}u^{\star}(w)
\end{align*}
Consider any $f \in F$. If $\mu(f)=\bar{\mu}(f)$, then
 \begin{align*}
     V_f(\mu(f))-\sum_{w \in \mu(f)}u^{\star}(w) = V_f(\bar{\mu}(f))-\sum_{w\in\bar{\mu}(f)}u^{\star}(w) \leq V_f(\mu^{\star}(f))-\sum_{w\in\mu^{\star}(f)}u^{\star}(w)
 \end{align*}
 If $\mu(f) \ne \bar{\mu}(f)$, then essential uniqueness implies $V_f(\mu(f))=V_f(\bar{\mu}(f))=0$. Hence, 
 \begin{align*}
     V_f(\mu(f))-\sum_{w\in\mu(f)}u^{\star}(w) = -\sum_{w\in\mu(f)}u^{\star}(w) \leq 0
 \end{align*}
 whereas individual rationality of $(\mu^{\star},u^{\star})$ implies 
 \begin{align*}
     V_f(\mu^{\star}(f))-\sum_{w\in\mu^{\star}(f)}u^{\star}(w)\geq 0
 \end{align*}
 Thus, in either case, 
 \begin{align*}
     V_f(\mu(f))-\sum_{w \in \mu(f)}u^{\star}(w) \leq V_f(\mu^{\star}(f))-\sum_{w\in\mu^{\star}(f)}u^{\star}(w)
 \end{align*}
By Lemma \ref{lem:domchar}, no outcome in $\mathcal{I}_\mu$ dominates $(\mu^\star,u^{\star})$. Since this is true for every $\mu \in \mathcal{M}(\mathcal{V})$,  no outcome in $\mathcal{V}$ dominates $(\mu^{\star},u^{\star})$:
\begin{align*}
    (\mu,u) \nrightarrow (\mu^\star,u^\star)~\qquad~\forall~(\mu,u) \in \mathcal{V}
\end{align*}
Since $(\mu^{\star},u^{\star})\notin \mathcal{V}$, this contradicts external stability. Hence, every efficient matching belongs to $\mathcal{M}(\mathcal{V})$. Using (\ref{eq:surplus}), we conclude that $\mathcal{M}(\mathcal{V})=\mathcal{M}^{\rm eff}$.

Finally, we argue that $\mathcal{V} = \bigcup_{\mu \in \mathcal{M}^{\rm eff}}\mathcal{I}_{\mu}$ (if the economy is generic, this will show that $\mathcal{V}$ contains the set of efficient and IR outcomes, as is necessary for Theorem \ref{theo:firmrich}). We have already established that 
$\mathcal{V} \subset \bigcup_{\mu \in \mathcal{M}^{\rm eff}}\mathcal{I}_{\mu}$. Assume for contradiction that some efficient outcome $(\mu^\star,u^\star) \notin \mathcal{V}$. Then, by external stability, there is some efficient outcome $(\mu,u) \in \mathcal{V}$ and a firm $f$ such that $(\mu,u) \xrightarrow[]{~f~}(\mu^\star,u^\star)$:
\begin{align*}
    u(w) &\ge u^\star(w)~\qquad~\forall~w \in \mu(f) \\
    V_f(\mu(f)) - \sum_{w \in \mu(f)}u(w) &\ge V_f(\mu^\star(f)) - \sum_{w \in \mu^\star(f)}u^\star(w)
\end{align*}
with one inequality strict. Since $\mathcal{M}^{\rm eff}$ is essentially unique, we get $V_f(\mu(f))=V_f(\mu^\star(f))$. Using this and adding all the previous inequalities, we get
\begin{align*}
    \sum_{w \in \mu^\star(f)}u^\star(w) &> \sum_{w \in \mu(f)}u^\star(w)
\end{align*}
If $\mu(f)=\mu^\star(f)$, this is a contradiction. Else, by essential uniqueness, we have $V_f(\mu(f))=V_f(\mu^\star(f))=0$. But then, IR implies that $u^\star(w)=0$ for all $w \in \mu(f)$ and for all $w \in \mu^\star(f)$. This will again contradict the previous inequality.
$\hfill \blacksquare{}$

\section{Omitted Proofs of Lemmas}
\label{sec:omapp}

\noindent {\bf Proof of Lemma \ref{lem:vnmempty}:} Let $\mathcal{V} \subseteq \mathcal{I}$ be a candidate vNM stable set. We consider multiple steps. \\

    \noindent {\sc Step 1.} Consider a matching $\mu \notin \{\mu_1,\mu_2,\mu_3\}$. We know that if $\mu \notin \{\mu_1,\mu_2,\mu_3\}$ and $(\mu,u_{w_1},u_{w_2},u_{w_3}) \in \mathcal{I}$, then $u_{w_k}=0$ for all $k \in \{1,2,3\}$. Hence, the only outcome corresponding to $\mu$ which may belong to $\mathcal{V}$ is $(\mu,0,0,0)$. Consider $(\mu_1,x,y,0) \in \mathcal{I}$ where $x+y \le 1$ and $x,y \ge 0$. In this case, $(\mu_1,x,y,0) \xrightarrow[]{~f_1~}(\mu,0,0,0)$. A similar argument shows that $(\mu_2,0,x,y) \xrightarrow[]{~f_2~}(\mu,0,0,0)$ and $(\mu_3,x,0,y) \xrightarrow[]{~f_3~}(\mu,0,0,0)$. Hence, if $(\mu,0,0,0) \in \mathcal{V}$ for some $\mu \notin \{\mu_1,\mu_2,\mu_3\}$, internal stability requires that no outcomes involving $\mu_1,\mu_2,\mu_3$ can be in $\mathcal{V}$. Hence, external stability requires that some $(\mu,0,0,0)$, where $\mu \notin \{\mu_1,\mu_2,\mu_3)$, must dominate $(\mu_1,x,y,0)$, which is not possible. Thus, $\mathcal{V}$ only contains outcomes corresponding to efficient matchings $\mu_1,\mu_2,\mu_3$. \\

    \noindent {\sc Step 2.} The outcome $(\mu_1,1,0,0)$ cannot be dominated by $(\mu_3,x,0,y)$. This is because if $(\mu_3,x,0,y)$ were to dominate $(\mu_1,1,0,0)$, then $x=1$ and hence, $y=0$, So, $(\mu_3,1,0,0)$ needs to dominate $(\mu_1,1,0,0)$, which is not possible because all firms get zero payoff in both the outcomes. Hence, it is not possible $\mathcal{V}=\mathcal{I}_{\mu_3}$. A similar argument extends this conclusion: it is not possible that $\mathcal{V}=\mathcal{I}_{\mu_i}$ for any $i \in \{1,2,3\}$. By Observation \ref{obs:is1}, $\mathcal{V} \nsubseteq \mathcal{I}_{\mu_i}$ for any $i \in \{1,2,3\}$. \\

    \noindent {\sc Step 3.} So, $\mathcal{V}$ must contain an outcome each from $\mathcal{I}_{\mu_i}$ for at least two values of $i$. Consider, without loss of generality, two outcomes $a_1 \equiv (\mu_1,x,y,0) \in \mathcal{I}$ and $a_2 \equiv (\mu_2,0,y',z') \in \mathcal{I}$. Note that if $y > y'$, then $a_1 \rightarrow a_2$ and if $y' > y$, then $a_2 \rightarrow a_1$. So, for internal stability, $y=y'$, i.e., $a_2 = (\mu_2,0,y,z')$. But if $z' > 0$, we have $a_2 \rightarrow a_1$. Hence, internal stability forces $z'=0$. Similarly, $x=0$. So, $a_1=(\mu_1,0,y,0)$ and $a_2=(\mu_2,0,y,0)$. If $y < 1$, then firm $1$ gets positive payoff in $a_1$ (and firm $2$ gets positive payoff in $a_2$), and $a_1 \rightarrow a_2$ in that case. So, internal stability forces $y=1$. Hence, for internal stability, the only outcomes in $\mathcal{I}_{\mu_1}$ and $\mathcal{I}_{\mu_2}$ that can be in $\mathcal{V}$ are $(\mu_1,0,1,0)$ and $(\mu_2,0,1,0)$. 

    Now, consider $a_3 \equiv (\mu_3,x'',0,z'')$. Can $a_3 \in \mathcal{V}$? Since $a_1 \in \mathcal{V}$, we argue that $a_3 \notin \mathcal{V}$ due to internal stability. This is because $a_3 \rightarrow a_1$ (since firm 3, and workers $1$ and $3$ get zero payoff in $a_1$, and at least one of them must get positive payoff in $a_3$). Hence $a_3 \notin \mathcal{V}$ for any values of $1 \ge x'' \ge 0$ and $1 \ge z'' \ge 0$. So, the only kind of outcomes in $\mathcal{V}$ has one of the matchings in $\{\mu_1,\mu_2,\mu_3\}$ and assigns the total surplus $1$ to one of the workers matched. 
    
    But then, consider $\hat{a}_3 \equiv (\mu_3,\frac{1}{2},0,\frac{1}{2})$. For external stability, either $a_1$ or $a_2$ must dominate $\hat{a}_3$. But $a_1$ cannot dominate $\hat{a}_3$ because worker $1$ gets positive payoff in $\hat{a}_3$ but gets zero in $a_1$. Similarly, $a_2$ cannot dominate $\hat{a}_3$ because worker $3$ gets positive payoff in $\hat{a}_3$ but gets zero in $a_2$. By Observation \ref{obs:is1}, no outcome involving $\mu_3$ can dominate $\hat{a}_3$. This implies that $\mathcal{V}$ fails to satisfy external stability.
    $\hfill \blacksquare{}$

    \medskip

    \noindent {\bf Proof of Lemma \ref{lem:order}:}
        We do the proof in two steps. \\

    \noindent {\sc Step 1.} We show that there exists some firm $f$ such that $\mu(f) \ne \mu^\star(f)$ and 
    \begin{align}\label{eq:fblock1}
        V_f(\mu^\star(f)) > V_f(\mu(f)) - \sum_{w \in \mu(f)}u(w) + \sum_{w \in \mu^\star(f)}u(w)
    \end{align}
    If this is not true, the reverse inequality holds for all firms whose matchings differ in $\mu$ and $\mu^\star$. Adding them together along with the firms whose matchings coincide (note the inequality (\ref{eq:fblock1}) is an equality if $\mu(f)=\mu^\star(f)$) gives
    \begin{align*}
        \sum_{f \in F}  V_f(\mu^\star(f)) &\le \sum_{f \in F} V_f(\mu(f))- \sum_{f \in F} \sum_{w \in \mu(f)}u(w) + \sum_{f \in F} \sum_{w \in \mu^\star(f)}u(w) \\
        &= \sum_{f \in F} V_f(\mu(f))- \sum_{w \in W}u(w) + \sum_{f \in F} \sum_{w \in \mu^\star(f)}u(w) \\
        &\le \sum_{f \in F} V_f(\mu(f))- \sum_{w \in W}u(w) +  \sum_{w \in W}u(w) \\
        &= \sum_{f \in F} V_f(\mu(f)),
    \end{align*}
    where the first equality follows from the fact that unmatched workers in $(\mu,u)$ must get zero payoff (IR) and the second inequality follows because $u(w) \ge 0$ for all $w \in W$ (again IR). But this contradicts the fact that $\mu$ is not an efficient matching and $\mu^\star$ is efficient. Hence, there exists a firm $f$ such that $\mu(f) \ne \mu^\star(f)$ and Inequality (\ref{eq:fblock1}) holds. Let 
    \begin{align}\label{eq:fblock2}
        \Delta := V_f(\mu^\star(f)) - V_f(\mu(f)) + \sum_{w \in \mu(f)}u(w) - \sum_{w \in \mu^\star(f)}u(w)
    \end{align}
    By Inequality (\ref{eq:fblock1}), we have $\Delta > 0$. \\

    \noindent {\sc Step 2.} We construct $u^\star$ such that $(\mu^\star,u^\star) \in \mathcal{I}_{\mu^\star}$ and $(\mu^\star,u^\star) \xrightarrow[]{~f~}(\mu,u)$, where $f$ is the identified firm in Step 1. The $u^\star$ is defined as follows:
    \begin{align*}
        u^\star(w) = 
        \begin{cases}
            u(w) + \frac{\Delta}{|\mu^\star(f)|} & \textrm{if}~w \in \mu^\star(f) \\
            0 & \textrm{otherwise}
        \end{cases}
    \end{align*}
    Since $u(w) \ge 0$, we have $u^\star(w) \ge 0$. Further, $u^\star(w) > u(w)$ for all $w \in \mu^\star(f)$. The utility of every firm $f' \ne f$ from the outcome $(\mu^\star,u^\star)$ is $V_{f'}(\mu^\star(f'))$ (since $u^\star(w) = 0$ for all $w \in \mu^\star(f')$), which is non-negative.\footnote{Even though $V_{f'}$ may take negative values, in an efficient matching $\mu^\star$, we must have $V_{f'}(\mu^\star(f')) \ge 0$ -- else not matching any worker to $f'$ improves aggregate surplus.} Finally, the utility of firm $f$ from outcome $(\mu^\star,u^\star)$ is
    \begin{align}\label{eq:equal1}
        V_f(\mu^\star(f)) - \sum_{w \in \mu^\star(f)}u^\star(w) = V_f(\mu^\star(f)) - \sum_{w \in \mu^\star(f)}u(w) - \Delta = V_f(\mu(f)) - \sum_{w \in \mu(f)}u(w)
    \end{align}
    where the last equality follows from (\ref{eq:fblock2}). Hence, the payoff of firm $f$ is the same in the two outcomes: $(\mu,u)$ and $(\mu^\star,u^\star)$. Since $(\mu,u) \in \mathcal{I}$, we thus conclude that $(\mu^\star,u^\star) \in \mathcal{I}$. Further, since each worker in $\mu^\star(f)$ is better off in $(\mu^\star,u^\star)$ than $(\mu,u)$, we conclude that $(\mu^\star,u^\star) \xrightarrow[]{~f~}(\mu,u)$.
    $\hfill \blacksquare{}$

    \medskip

    \noindent {\bf Proof of Lemma \ref{lem:domchar}.} $(2) \Rightarrow (1)$. Suppose for some $(\mu,u)\in \mathcal{I}_{\mu}$, we have $(\mu,u)~\xrightarrow[]{~f~}~(\mu',u')$ for some firm $f$. Then, 
        \begin{align*}
            u(w) &\ge u'(w)~\qquad~\forall~w \in \mu(f) \\
            V_f(\mu(f))-\sum_{w\in \mu(f)}u(w) &\ge V_f(\mu'(f))-\sum_{w\in \mu'(f)}u'(w)
        \end{align*}
        with one inequality strict. Adding these inequalities contradicts (\ref{eq:dcc1}) for $f$. Hence $(2) \Rightarrow (1)$. \\
        
        \noindent $(1) \Rightarrow (2)$. Suppose (\ref{eq:dcc1}) does not hold for firm $f$. That is, 
        \begin{align}
                V_f(\mu(f))-\sum_{w\in \mu(f)}u'(w) > V_f(\mu'(f))-\sum_{w\in \mu'(f)}u'(w) \label{eq:dcc11}
        \end{align}
        Let $(\mu,\hat{u})\in \mathcal{I}_{\mu}$ (such an outcome exists since $\mu$ is IR-feasible). Define the outcome $(\mu,u)$ as follows
        \begin{align*}
        u(w) = 
        \begin{cases}
            u'(w)  & \textrm{if}~w \in \mu(f) \\
            \hat{u}(w) & \textrm{otherwise}
        \end{cases}
    \end{align*}
    
    We first show that $(\mu,u) \in \mathcal{I}_{\mu}$. Since $(\mu',u'),(\mu,\hat{u})\in \mathcal{I}$, $u(w)\geq 0$ for all $w\in W$ and for all $f' \ne f$, 
    \begin{align*}
        0 \le V_{f'}(\mu(f')) - \sum_{w \in \mu(f')}\hat{u}(w) = V_{f'}(\mu(f')) - \sum_{w \in \mu(f')}u(w)
    \end{align*}
    where the inequality follows from $(\mu,\hat{u}) \in \mathcal{I}$ and the equality follows from definition of $u$. For firm $f$, 
    \begin{align*}
        V_f(\mu(f))-\sum_{w\in\mu(f)}u(w)=V_f(\mu(f))-\sum_{w\in\mu(f)}u'(w) > V_f(\mu'(f))-\sum_{w\in\mu'(f)}u'(w) \ge 0
    \end{align*}
    where the strict inequality is due to (\ref{eq:dcc11}) and last inequality is true because $(\mu',u') \in \mathcal{I}$. Hence, $(\mu,u)\in \mathcal{I}_{\mu}$. 
    
    Further, by construction of $u$, $u(w)=u'(w)$ for all $w\in \mu(f)$ and (\ref{eq:dcc11}) ensures that 
    \begin{align*}
        V_f(\mu(f))-\sum_{w\in\mu(f)}u(w)>V_f(\mu'(f))-\sum_{w\in\mu'(f)}u'(w)
    \end{align*}
    Thus,  $(\mu,u)\xrightarrow[]{~f~} (\mu',u')$. Hence, $(1) \Rightarrow (2)$.
   $\hfill \blacksquare{}$

   \medskip

   \noindent {\bf Proof of Lemma \ref{lem:flow}.} Suppose there exists a flow $\phi$ in the directed graph. Fix $w \in W$. Suppose there is an incoming edge from $\hat{f}$ (i.e., $f \equiv \mu(w)$) and an outgoing edge to $f' \equiv \mu'(w)$ at $w$, then flow balancing will imply $\phi(w,f')=\phi(\hat{f},w)$, and then set $u'(w)=\phi(w,f')=\phi(\hat{f},w)$.

    We show that this $u'$ satisfies (\ref{eq:es}) and (\ref{eq:ir}). To see this, pick $f \in F$. The inflow to vertex $f$ is equal to the flow from $f$ to $\hat{f}$ (since $b(f)=0$ and flow balancing must hold), and inflow (and outflow) is equal to
    \begin{align*}
        \sum_{(w,f) \in E}\phi(w,f) = \sum_{w \in \mu'(f)}u'(w) \le c(f,\hat{f}) = V_f(\mu'(f))
    \end{align*}
    where the inequality follows because flow satisfies capacity constraint. Hence, (\ref{eq:ir}) holds.

    Now, consider vertex $\hat{f}$ and flow balancing at this vertex. Total inflow is equal to sum of inflows from $f$ and from $v_0$. Flow from $f$ to $\hat{f}$ is equal to the inflow to vertex $f$, which in turn is equal to $\sum_{w \in \mu'(f)}u'(w)$. Hence, total inflow to $\hat{f}$ is equal to
    \begin{align*}
        \sum_{w \in \mu'(f)}u'(w) + \phi(v_0,\hat{f})
    \end{align*}
    Total outflow from $\hat{f}$ is 
    \begin{align*}
        \sum_{(\hat{f},w) \in E}\phi(\hat{f},w) = \sum_{w \in \mu(f)}\phi(\hat{f},w)=\sum_{w \in \mu(f)}u'(w)
    \end{align*}
    Using the fact that $b(\hat{f})=V_f(\mu'(f))-V_f(\mu(f))$, flow balancing at $\hat{f}$ gives us
    \begin{align*}
        \sum_{w \in \mu'(f)}u'(w) + \phi(v_0,\hat{f}) - \sum_{w \in \mu(f)}u'(w) = V_f(\mu'(f)) - V_f(\mu(f))
    \end{align*}
    Since $\phi(v_0,\hat{f}) \ge 0$, we get 
    \begin{align*}
        \sum_{w \in \mu'(f)}u'(w) - \sum_{w \in \mu(f)}u'(w) \le V_f(\mu'(f)) - V_f(\mu(f))
    \end{align*}
    Hence, (\ref{eq:es}) holds for firm $f$.

    The converse proof follows similar arguments. Suppose $u'$ is a feasible solution to (\ref{eq:es}) and (\ref{eq:ir}). Then, we can construct a flow in $\mathcal{G}$ from it. For every $w \in W$, since $\mu'(w) \ne \emptyset$ and $\mu(w) \ne \emptyset$, we can set $\phi(w,\mu'(w))=\phi(\hat{f},w):=u'(w)$, where $\mu(w)=f$. This balances flow at $w$. Now, at any $f \in F$, let
    \begin{align*}
        \phi(f,\hat{f}) := \sum_{w \in \mu'(f)}u'(w)= \sum_{w \in \mu'(f)}\phi(w,f)
    \end{align*}
    This balances flow at $f$. Further, since $u'$ satisfies (\ref{eq:ir}), this means
    \begin{align*}
        \phi(f,\hat{f})=\sum_{w \in \mu'(f)}u'(w) \le V_f(\mu'(f)) = c(f,\hat{f})
    \end{align*}
    Hence, capacity constraint on edge $(f,\hat{f})$ holds.\footnote{Capacity of other edges is greater than $V(\mu')$ and $u'(w) \le V_{\mu'(w)}(\mu'(\mu'(w))) \le V(\mu')$ implies capacity constraints of edges $(w,\mu'(w))$ and $(\hat{f},w)$, where $\mu(w)=f$, also hold.}
    
    Now to define the flow $(v_0,\hat{f})$, let
    \begin{align*}
        \phi(v_0,\hat{f}) := \big[V_f(\mu'(f)) - V_f(\mu(f))\big] - \big[ \sum_{w \in \mu'(f)}u'(w) - \sum_{w \in \mu(f)}u'(w)\big]
    \end{align*}
    Since $u'$ satisfies (\ref{eq:es}), this is non-negative. Also, 
    \begin{align*}
        \phi(v_0,\hat{f}) \le V_f(\mu'(f)) + \sum_{w \in \mu(f) \setminus \mu'(f)}u'(w) \le V_f(\mu'(f)) + \sum_{f' \ne f}V_{f'}(\mu'(f')) \le V(\mu') < c(v_0,\hat{f}),
    \end{align*}
    where the second inequality uses the fact that $(\mu',u')$ is an IR outcome and the last inequality uses the fact that the capacity of edge $(v_0,\hat{f})$ is sufficiently large. Hence, capacity constraint of edge $(v_0,\hat{f})$ holds for each $f \in F$.
    
    Next,
    \begin{align*}
        \phi(v_0,\hat{f}) = \big[V_f(\mu'(f)) - V_f(\mu(f))\big] - \big[ \sum_{w \in \mu'(f)}u'(w) - \sum_{w \in \mu(f)}u'(w)\big] &= b(\hat{f}) - [\phi(f,\hat{f}) - \sum_{w \in \mu(f)}\phi(\hat{f},w)] \\
        \Longrightarrow \phi(v_0,\hat{f}) + \phi(f,\hat{f}) - \sum_{w \in \mu(f)}\phi(\hat{f},w) &= b(\hat{f})
    \end{align*}
    This means flow balancing holds at $\hat{f}$.

    Finally, at vertex $v_0$,
    \begin{align*}
        \sum_{f \in F}\phi(v_0,\hat{f}) &= \sum_{f \in F}\big[V_f(\mu'(f)) - V_f(\mu(f))\big] - \sum_{f \in F}\big[ \sum_{w \in \mu'(f)}u'(w) - \sum_{w \in \mu(f)}u'(w)\big] \\
        &= \sum_{f \in F}\big[V_f(\mu'(f)) - V_f(\mu(f))\big] = -b(v_0)
    \end{align*}
    where the second equality follows because all workers are matched in $\mu'$ and $\mu$, and hence, 
    \begin{align*}
        \sum_{f \in F}\big[ \sum_{w \in \mu'(f)}u'(w) - \sum_{w \in \mu(f)}u'(w)\big] = 0
    \end{align*}
    Since there are no incoming edges to $v_0$, this shows that flow balancing holds at $v_0$ also.
    $\hfill \blacksquare{}$

   \section{A vNM stable set with only inefficient outcomes}
   \label{sec:exampleapp}

   \begin{example}[a vNM stable set with only inefficient outcomes]
    \label{ex: inefficient VNM stable set}
    {\rm Consider the following economy. There are three firms $F=\{f_1,f_2,f_3\}$ and three workers $W=\{w_1,w_2,w_3\}$. The surplus generated by each firm is summarized in Table \ref{tab: surplus(ineffiecient stable set)}.
    \begin{table}[!hbt]
        \centering
        \begin{tabular}{|c||c|c|c|c|c|c|c|}
        \hline
             & $\{w_1\}$ & $\{w_2\}$ & $\{w_3\}$ & $\{w_1,w_2\}$ & $\{w_2,w_3\}$ & $\{w_3,w_1\}$ & $\{w_1,w_2,w_3\}$ \\
             \hline 
        $V_{f_1}(\cdot)$ & -1 & -1 & -1 & 4 & -1 & 2 & -1 \\
        $V_{f_2}(\cdot)$ & 3 & 3 & 1 & -1 & -1 & -1 & -1 \\
        $V_{f_3}(\cdot)$ & -1 & -1 & 2 & -1 & 2 & -1 & -1 \\
        \hline
        \end{tabular}
        \caption{Surplus of firms and workers}
        \label{tab: surplus(ineffiecient stable set)}
    \end{table}

    There is a unique efficient matching $\mu^\star$ given by $\mu^\star(f_1)=\{w_1,w_2\}$, $\mu^\star(f_2)=\emptyset$, and $\mu^\star(f_3)=\{w_3\}$. The core is empty for this economy.

Consider the following three inefficient matchings:
$$\begin{array}{c@{\qquad}c@{\qquad}c@{\qquad}c}
	\mu_1: &
	\mu_1(f_1)=\{w_1,w_3\} &
	\mu_1(f_2)=\{w_2\} &
	\mu_1(f_3)=\emptyset \\[1mm]
	\mu_2: &
	\mu_2(f_1)=\{w_1,w_2\} &
	\mu_2(f_2)=\{w_3\} &
	\mu_2(f_3)=\emptyset \\[1mm]
	\mu_3: &
	\mu_3(f_1)=\emptyset &
	\mu_3(f_2)=\{w_1\} &
	\mu_3(f_3)=\{w_2,w_3\}
\end{array}$$

Let $\bar{u}$ be given by $\bar{u}(w_1)=\bar{u}(w_2)=2$ and $\bar{u}(w_3)=0$. For every $i\in \{1,2,3\}$, define the outcome  $a_i=(\mu_i,\bar u)$. At each of the outcomes $a_1$, $a_2$, and $a_3$, firms $f_1$ and $f_3$ receive payoff $0$, while firm $f_2$ receives payoff $1$. Moreover, workers $w_1$ and $w_2$ receive payoff $2$, while worker $w_3$ receives payoff $0$.  Now define the set of outcomes $$\mathcal{U}=\left\{ (\mu_1,u)\in\mathcal I_{\mu_1}: u(w_3)>0,\; u(w_1)+u(w_3)<2,\; u(w_2)<2 \right\}.$$ Let $\mathcal V=\mathcal{U}\cup\{a_1,a_2,a_3\}.$ By construction, $\mathcal{V}\subseteq \mathcal{I}$. We show that $\mathcal{V}$ is a vNM stable set.

\noindent \textbf{Internal stability:} By Observation \ref{obs:is1}, no two outcomes in $\mathcal{U}\cup \{a_1\}$ can dominate one another. Moreover, the payoff of each worker and firm is the  same at $a_1,a_2$ and $a_3$. Hence, none of these outcomes dominates another. It remains to compare outcomes in $\mathcal U$ with $a_2$ and $a_3$. Consider any $(\mu_1,u)\in\mathcal U$, and let $u(w_1)=x$, $u(w_2)=y$, and $u(w_3)=z$. By the definition of $\mathcal U$, $z>0$, $x+z<2$, and $y<2$. 
Hence, the payoffs of firms $f_1,f_2$ and $f_3$ at $(\mu_1,u)$ are  $2-x-z>0$, $3-y>1$ and $0$, respectively. 

Since $a_1$, $a_2$, and $a_3$ give the firms and workers the same payoff, if $(\mu_1,u)$ dominates either $a_2$ or $a_3$, it would also dominate $a_1$. This contradicts Observation \ref{obs:is1}. Hence, $(\mu_1,u)\nrightarrow a_2$ and $(\mu_1,u)\nrightarrow a_3$.

At  $a_2$ and $a_3$, firm $f_1$ receives payoff $0$, so neither outcome can dominate $(\mu_1,u)$ through $f_1$. At $a_3$, firm $f_2$ receives payoff $1$, whereas its payoff at $(\mu_1,u)$ is strictly greater than $1$. Hence, $a_3$ cannot dominate $(\mu_1,u)$ through $f_2$. At $a_2$, firm $f_2$ is matched to
$w_3$, but $w_3$ receives payoff $0$ at $a_2$ and $z>0$ at $(\mu_1,u)$. Thus, $a_2$ cannot dominate $(\mu_1,u)$ through $f_2$. At $a_2$ and $(\mu_1,u)$, $f_3$ is unmatched and receives payoff $0$ at both outcomes, so there is no strict improvement. At $a_3$, $f_3$ is matched to $\{w_2,w_3\}$, but $w_3$ receives $0$ at $a_3$ and $z>0$ at $(\mu_1,u)$. Hence, $a_2\nrightarrow(\mu_1,u)$ and $a_3\nrightarrow(\mu_1,u)$. Thus, no two outcomes in $\mathcal V$ dominate one another, and $\mathcal V$ satisfies internal stability.

\noindent \textbf{External stability:}  Consider any $(\mu,u)\in\mathcal I\setminus\mathcal V$.  Since $(\mu,u)$ is individually rational, $f_2$ can only be matched to $\emptyset$, $\{w_1\}$, $\{w_2\}$, or $\{w_3\}$. We consider these four cases.

\medskip
\noindent\textit{Case 1: $\mu(f_2)=\emptyset$.} \\
In this case the payoff of $f_2$ is zero. If $u(w_2)\leq2$, then $a_1\xrightarrow{f_2}(\mu,u)$. If $u(w_2)>2$, then $w_2$ cannot be unmatched. Moreover, since $(\mu,u)\in\mathcal{I}$ and $\mu(f_2)=\emptyset$,  we must have $\mu(f_1)=\{w_1,w_2\}$ ($\mu(f_3)=\{w_2,w_3\}$ is not possible since $V_{f_3}(\{w_2,w_3\})=2 < u(w_2)$, which will contradict IR for firm $f_3$). Since $V_{f_1}(\{w_1,w_2\})=4$ and $u(w_2)>2$, it follows that $u(w_1)<2$. Hence, $a_3\xrightarrow{f_2}(\mu,u)$.

\medskip
\noindent\textit{Case 2: $\mu(f_2)=\{w_1\}$.} \\
Since $(\mu,u)\in \mathcal{I}$, we must have  $\mu(f_1)=\emptyset$ and $u(w_2)\leq 2$. If $u(w_1)<2$, then $a_2\xrightarrow{f_1}(\mu,u)$. If $u(w_1)>2$, then the payoff of  $f_2$ is strictly less than $1$. Combining this with the fact that $u(w_2)\leq 2$,  $a_1\xrightarrow{f_2}(\mu,u)$. If $u(w_1)=2$, the payoff of $f_2$ is $1$. Hence, if $u(w_2)< 2$, then $a_1\xrightarrow{f_2}(\mu,u)$. However, if $u(w_2)=2$, by IR of $(\mu,u)$, $\mu(f_3)=\{w_2,w_3\}$ and $u(w_3)=0$. But, then $(\mu, u)=a_3$, contradicting  $(\mu,u)\notin \mathcal{V}$.

\medskip
\noindent\textit{Case 3: $\mu(f_2)=\{w_2\}$.}\\
Since $(\mu,u)\in \mathcal{I}$, in this case either $\mu(f_1)=\emptyset$ or $\mu(f_1)=\{w_1,w_3\}$. Suppose $\mu(f_1)=\emptyset$. Then, the payoff of $f_1$ is zero. Since $w_1$ cannot be matched to $f_3$ at an IR outcome, $w_1$ is unmatched and hence $u(w_1)=0$. If $u(w_2)< 2$, then $a_2\xrightarrow{f_1}(\mu,u)$. If $u(w_2)\geq 2$, then the payoff of $f_2$ is weakly less than $1$. Hence, $a_3\xrightarrow{f_2}(\mu,u)$.  Now suppose $\mu(f_1)=\{w_1,w_3\}$. By IR of $(\mu,u)$, $u(w_1)+u(w_3)\leq2$. If $u(w_2)>2$, then $a_3\xrightarrow{f_2}(\mu,u)$. Suppose $u(w_2)=2$. If $u(w_1)<2$, then $a_3\xrightarrow{f_2}(\mu,u)$. If $u(w_1)=2$, then IR implies $u(w_3)=0$, so $(\mu,u)=a_1$, contradicting $(\mu,u) \notin \mathcal{V}$. Finally, suppose $u(w_2)<2$. If $u(w_3)=0$, then $a_3\xrightarrow{f_3}(\mu,u)$. If $u(w_3)>0$, then, since $(\mu,u)\notin\mathcal U$, we must have $u(w_1)+u(w_3)=2$. Hence, $a_2\xrightarrow{f_1}(\mu,u)$.

\medskip
\noindent\textit{Case 4: $\mu(f_2)=\{w_3\}$.}\\
Since $(\mu,u)\in \mathcal{I}$, in this case $\mu(f_3)=\emptyset$ and either $\mu(f_1)=\emptyset$ or $\mu(f_1)=\{w_1,w_2\}$. Moreover, $u(w_3)\leq 1$ and hence the payoff of $f_2$ is at most $1$. If $\mu(f_1)=\emptyset$, then  $a_2\xrightarrow{f_1}(\mu,u)$. Suppose $\mu(f_1)=\{w_1,w_2\}$. Since $V_{f_1}(\{w_1,w_2\})=4$, IR implies $u(w_1)+u(w_2) \leq 4$. If $u(w_1)< 2$, then $a_3\xrightarrow{f_2}(\mu,u)$ and if $u(w_2)< 2$, then $a_1\xrightarrow{f_2}(\mu,u)$. The only remaining case is $u(w_1)=u(w_2)=2$. If $u(w_3)>0$, then $a_1\xrightarrow{f_2}(\mu,u)$ and $a_3\xrightarrow{f_2}(\mu,u)$. If $u(w_3)=0$, then $(\mu, u)=a_2$, contradicting $(\mu,u)\notin \mathcal{V}$.

Thus, every outcome in $\mathcal I\setminus\mathcal V$ is dominated by an outcome in $\mathcal{V}$. Hence, $\mathcal V$ satisfies
external stability.}
\end{example}

\end{document}